\documentclass[a4paper,11pt]{article}

\usepackage{amsmath,amscd,amssymb}
\usepackage[noadjust]{cite}
\usepackage{graphicx}
\usepackage{url}
\usepackage{verbatim} 
\usepackage{multirow}
\usepackage{algorithm}
\usepackage[noend]{algorithmic}
\usepackage{xspace}
\usepackage{epsfig}
\usepackage{Definitions}
\usepackage{empheq}
\usepackage{multirow,amscd,amssymb}
\usepackage{colortbl}
\usepackage{paralist}
\usepackage{subfig}
\usepackage{tabularx}
\usepackage{hhline}
\usepackage{footnote}

\usepackage[algo2e,vlined,linesnumbered,ruled]{algorithm2e}

\begin{document}
%
\title{Uncovering the Spatiotemporal Patterns of Collective Social Activity}


\author{Martin Jankowiak\thanks{
Center for Urban Science and Progress, New York University, jankowiak@gmail.com}
\and
Manuel Gomez-Rodriguez\thanks{Max Planck Institute for Software Systems, manuelgr@mpi-sws.org}}
\date{}

\maketitle

\begin{abstract}
Social media users and microbloggers post about a wide variety of (off-line) collective social activities as they participate in them, ranging from concerts and sporting events to political rallies and civil protests.
In this context, people who take part in the same collective social activity often post closely related content from nearby locations at similar 
times, resulting in distinctive spatiotemporal patterns. Can we automatically detect these patterns and thus provide insights into the associated
activities?
In this paper, we propose a modeling framework for clustering streaming spatiotemporal data, the Spatial Dirichlet Hawkes Process (SDHP), 
which allows us to automatically uncover a wide variety of spatiotemporal patterns of collective social activity from geolocated online traces. 
Moreover, we develop an efficient, online in\-fe\-rence algorithm based on Sequential Monte Carlo that scales to millions of geolocated posts.
Experiments on synthetic data and real data gathered from Twitter show that our framework can recover a wide variety of meaningful social activity
patterns in terms of both content and spatiotemporal dynamics, that it yields interesting insights about these patterns, and that 
it can be used to estimate the location from where a tweet was posted.
\end{abstract}

\section{Introduction}
\label{sec:intro}
With the widespread adoption of smartphones, the use of social media has become increasingly common at 
social gatherings and a wide variety of (off-line) collective social activities.
For example, a supporter at a political rally may post about a politician on Twitter; 
a music fan may upload a video of a concert
or an art enthusiast may upload a photo of a painting to Instagram.
In this context, people taking part in the same collective social activity---a political rally, a concert, or an art exhibition---may use their smartphones 
to post closely related content from nearby locations at similar times.
Therefore, one would expect that (sufficiently large) collective social activities will induce distinctive spatiotemporal patterns in social media and
online social networking sites.

\begin{figure}[t]
\centering
\includegraphics[width=3.6in]{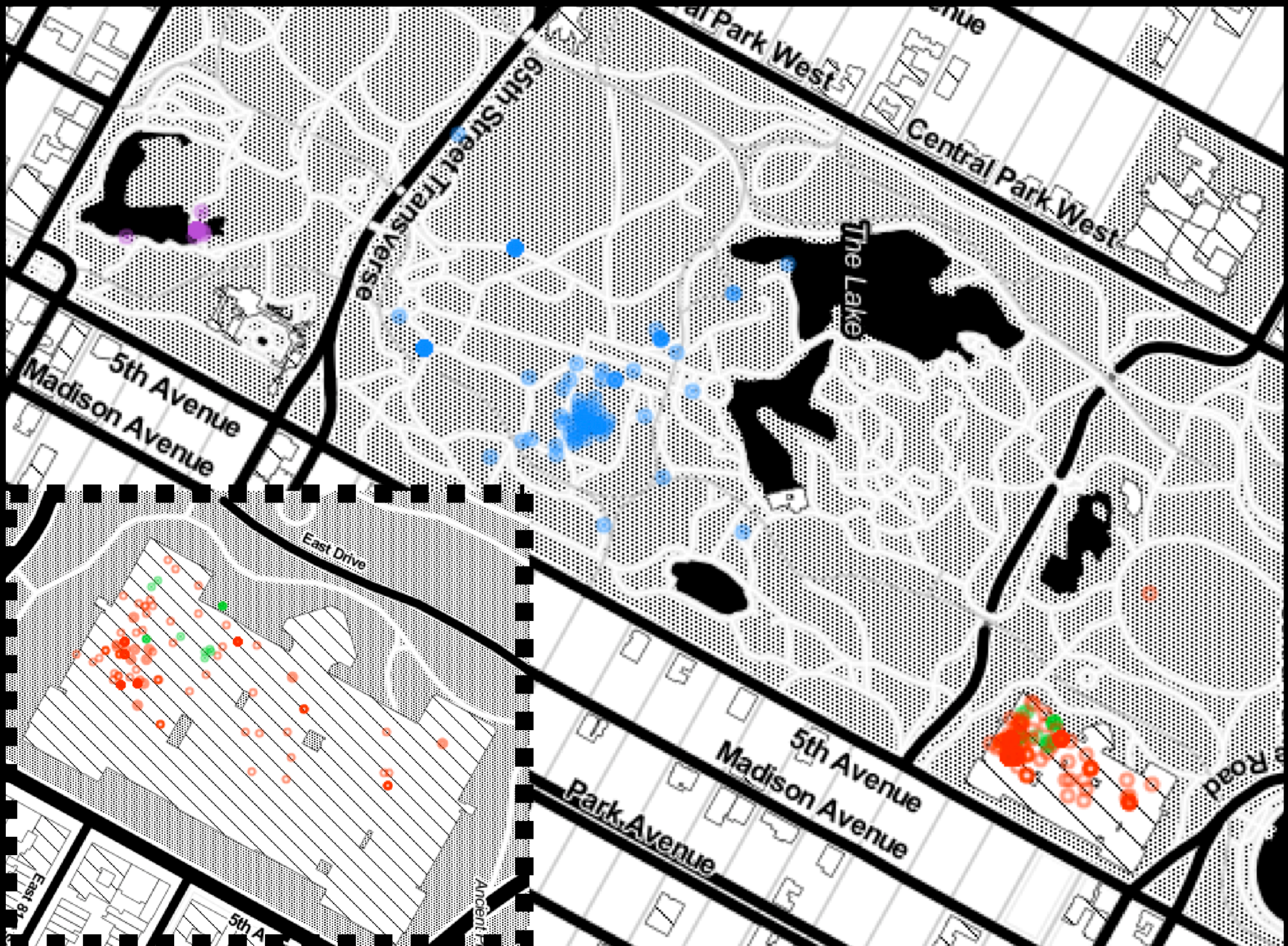}
\caption{Tweet locations for four spatiotemporal patterns inferred by the SDHP. The underlying collective social activities correspond to a typical visit to the Metropolitan Museum of Art 
(the Met; in red), journalists tweeting at a press preview at the Met (in green), a typical visit to Gapstow Bridge (in purple), and musical concerts held
at Rumsey Playfield and Naumburg Bandshell (in blue). Each circle corresponds to a tweet.}
\label{fig:central_park}
\end{figure}

In this paper, our goal is to develop an algorithmic framework to automatically identify, track and analyze these spatiotemporal patterns from a never-ending 
stream of timestamped geolocated content, be it text, photos or videos. 
Such a framework will enable us to detect and provide insights into the associated collective social activities as they unfold over time.
However, there are several challenges we need to address, which we illustrate next using a real world example from Central Park in New York, 
shown in Figure~\ref{fig:central_park}:

\emph{--- Uncertain location, time and content.} In most cases, there is uncertainty in the location, time, and content of a given
collective social activity and its associated posts in social media. For example, people attending a series of concerts at Rumsey Playfield (in blue)
posted at random times before, during and after one or more of the concerts from random locations near the stage. Therefore, we need to consider probabilistic 
models to capture these uncertainties so that the corresponding spatiotemporal and content distributions depend on the particular collective 
social activity.

\emph{--- Spatiotemporal heterogeneity.} Collective social activities span multiple spatial and temporal scales. For example, the concerts 
at Rumsey Playfield take place in a relatively large area over the entire summer, while visits to Gapstow bridge (in purple) 
span a few meters and occur throughout the year. Therefore, artificially discretizing the time and spatial axes into bins may 
introduce tuning parameters, like the bin size, which are not easy to choose optimally under such heterogeneity. 

\emph{--- Interleaved spatiotemporal patterns.} Collective social activities are often interleaved in terms of location, time and content. 
For example, typical visits to the Metropolitan Museum of Art (in red) occur year-round; however, visits to a press preview
within the same museum (in green) may span only a few hours but overlap with a typical visit both in terms of time and space.

To overcome the above challenges, we first present a novel continuous spatiotemporal probabilistic model, the Spatial Dirichlet Hawkes Process (SDHP), 
for clustering streaming geolocated text data.\footnote{While the same framework could also be used to model streaming geolocated videos or images by
swapping out the content model, here we focus on textual content, which allows us to do detailed experiments with Twitter data.}
In comparison with the Dirichlet Hawkes Process (DHP)~\cite{du2015dirichlet}, a recently proposed continuous-time model for clustering document streams, the 
key innovation of our work is to explicitly model the spatial information present in streaming geolocated data. 
As a consequence, our design is especially well suited to capturing and disentangling the spatiotemporal dynamics of a wide range of collective social activities.
We then exploit temporal dependencies in the observed data and conjugacies 
in our probabilistic model 
to develop an efficient, online inference algorithm based on Sequential Monte Carlo that scales to millions 
of spatiotemporal posts.

Finally, we validate our modeling framework using both synthetic data and real data gathered from Twitter, which comprises $\sim$$3.75$ million tweets posted by
users in New York City during a one year period from January 2013 to December 2013. Our results show that:
\begin{itemize}\setlength\itemsep{-.2em}
\item[1.] Our inference algorithm can accurately recover the model parameters and assign each geolocated post to the true spatiotemporal pattern given
only its spatiotemporal coordinates and content. Moreover, the sparser the content, the more our model benefits from 
utilizing spatial information.
\item[2.] We can recover meaningful real world spatiotemporal patterns across a diverse range of temporal and spatial scales.
These patterns correspond to a wide variety of collective social activities, 
from a series of concerts held throughout the summer in a park to tourist visits to a small bridge year-round.
\item[3.] Our model can be used to accurately estimate the location from where a user posted while taking part in a collective social activity, 
achieving comparable or better predictive performance than alternatives at a comparable or lower computational cost.
\end{itemize}

\section{Related Work}
\label{sec:related}
There is an extensive literature of models for clustering streaming data~\cite{du2015dirichlet,ahmed2011online,ahmed2011unified,ahmed2008dynamic,blei2011distance,diao2014recurrent}, 
typically motivated by applications in topic modeling. 
However, most of these models do not take spatial information into account, discretize the time axis into bins (thus introducing additional parameters which are difficult
to set), and ignore the mutual excitation between events, which has been observed in social activity data~\cite{farajtabar2014activity}.
Very recently, Du et al.~\cite{du2015dirichlet} proposed the Dirichlet Hawkes Process (DHP), a continuous-time model for clustering document streams such 
as news articles that addresses some of these limitations.
However, the DHP does not consider the spatial information present in streaming geolocated data and, as a consequence, our framework compares favorably 
at recovering spatiotemporal patterns of collective social activity.

In addition our work is related to the extensive literature on mining and modeling mobility patterns~\cite{cho2011friendship,wang2011human,becker2013human,lane2014connecting}. 
There are two fundamental differences between our work and this line of work:
(i) they typically focus on modeling and predicting the movements of groups or individuals; in contrast, we are interested in identifying and tracking collective 
social activities spanning multiple spatial and temporal scales; and
(ii) they only consider time and location; however, many geolocated social media data also contain content, which often reveals key aspects of the underlying
social activity.

Finally, there is a recent line of work on detecting events from the Twitter stream~\cite{abdelhaq2013eventweet,walther2013geo,weng2011event,chierichetti2014event,ihler2006adaptive}.
However, they share at least one of the following limitations:
(i) they only leverage one modality of data: spatial, temporal or content information;
(ii) they discretize the time and/or space axis into bins, introducing additional parameters that are difficult to set;
(iii) they do not provide insights into the underlying spatiotemporal dynamics of the events they detect; and
(iv) they focus on events taking place on similar temporal scales.

\section{Preliminaries}
\label{sec:preliminaries}
In this section, we review the three major building blocks of the Spatial Dirichlet Hawkes Process (SDHP): the Dirichlet Process~\cite{antoniak1974mixtures}, the 
Hawkes Process~\cite{hawkes1971spectra}, and the Normal distribution under the Normal-Gamma prior~\cite{murphy2007}.

\subsection{Dirichlet Process}
The Dirichlet process $DP(\beta, G_0)$ is a well known Bayesian nonparametric prior, parametrized by a concentration parameter
$\beta > 0$ and a base distribution $G_0(\theta)$ over a given space $\theta \in \Theta$. A sample $G \sim DP(\beta, G_0)$ drawn 
from a DP is a discrete distribution. 
One can generate samples from a DP using a generative process called the Chinese Restaurant Process (CRP). 
The CRP assumes a Chinese restaurant with an infinite number of tables, each corresponding to a cluster, so that whenever
a new customer arrives, she can either choose an existing table $k$ with $m_k$ seated customers or sit at an empty table. More specifically, one can obtain samples from the DP as follows:
\begin{itemize}
\setlength\itemsep{-0.3em}
\item[1.] Assign the first customer to an empty table and draw its parameter $\theta_1$ from $G_0$.
\item[2.] For $n > 1$:
\vspace{-.2cm}
\begin{itemize}\setlength\itemsep{-0.1em}
\item[--] With probability $\frac{\beta}{\beta+n-1}$ assign customer $n$ to an empty table and draw 
$\theta_n$ from $G_0$.
\item[--] With probability $\frac{m_k}{\beta+n-1}$ assign customer $n$ to the non-empty table $k$ and
reuse $\theta_k$ for $\theta_n$.
\end{itemize}
\end{itemize}
Since a new cluster can be created at each step with non-zero probability, the number of clusters is potentially infinite and 
thus the process can adapt to increasing complexity in the data.

\subsection{Hawkes Process}
A Hawkes process is a type of temporal point process \cite{AalBorGje08}, which is a stochastic process whose realization consists 
of a sequence of discrete events localized in time, $\Hcal = \cbr{t_i \in \RR^{+}\,|\,t_i < t_{i+1}}$. 
A temporal point process can also be re\-pre\-sen\-ted as a counting process via $N(t)$, which is the number of events up to time $t$. 
Moreover, we can characterize the counting process using the conditional intensity function $\lambda^*(t) $, which is the conditional probability of observing an event
in an infinitesimal time window $[t, t+dt)$ given the history $\Hcal(t) = \{ t_i \in \Hcal \,\vert\, t_i < t \}$, i.e.
\begin{equation}
\EE[dN(t)|\Hcal(t)] = \PP\cbr{\text{event in $[t, t+dt) | \Hcal(t)$}} = \lambda^*(t) dt \nonumber
\end{equation}
where $dN(t) \in \{0, 1\}$. 
In the case of Hawkes processes, the intensity function takes the form
\begin{equation}\label{eq:hawkes}
\lambda^*(t) = \lambda_{0} +\! \!\!\sum_{t_i\in \Hcal(t)} \!\!\gamma(t,t_i)
\end{equation}
where $\lambda_{0}$ is the base intensity and $\gamma(\cdot)$ is the triggering kernel.
Hawkes processes have been increasingly used to model social activity~\cite{farajtabar2014activity,zhao2015seismic}, since they are able to capture mutual excitation between events and thus open up the possibility of modeling bursts of rapidly occurring events separated by long periods of inactivity~\cite{barabasi05human}.

\subsection{Normal distributions with unknown mean and variance}
A conjugate prior for normally distributed data with unknown mean and variance, $x \sim \mathcal{N}( \cdot | \mu, \sigma)$, 
is the normal-gamma prior given by
\begin{align}
 \mathcal{N}( \mu | \mu_0, (\kappa_0 \lambda)^{-1}) \; {\rm Ga}(\lambda | \alpha_0, \beta_0)
\end{align}
where $\lambda \equiv 1/\sigma^2$.

\section{Proposed Model}
\label{sec:formulation}
In this section we build on the above and formulate our model for clustering streaming spatiotemporal data, 
the Spatial Dirichlet Hawkes Process (SDHP).
We first describe the geolocated posts it is designed to model, then elaborate on the different pieces
used to model the time, content and location of each post, and finally introduce a generative process view of the model.


\subsection{Geolocated post data}
Given an online social network or media site, we represent each geolocated post created by a user as a tuple $(t, \mathbf{d}, \mathbf{r}, s)$,
where $t$ is the time when the post was posted, $\mathbf{d}$ is the post content, $\mathbf{r}$ is the location
(e.g.~latitude, longitude) from where the user posted the post, and $s$ is the associated spatiotemporal 
pattern, which is latent. Throughout we will use $N$ to denote the number of posts in a given dataset $\mathcal{D}$.
%


\subsection{Intensity functions}
The time $t$ associated to each geolocated post is drawn from a Hawkes process with intensity given by
\begin{equation}
\lambda(t) = \lambda_0+\sum_{s} \lambda_{s}(t)
\end{equation}
where the constant $\lambda_0$ is the base intensity and 
where the pattern-specific intensity is given by
\begin{equation}
\lambda_{s}(t_n) = \sum_{i=1}^{n-1}\gamma_{s_i}(t_n,t_i) \II(s_i = s)
\end{equation}
and $\gamma_{s}(t, t_i) = \alpha_{s} e^{-(t-t_i)/\tau_{s}}$ is an exponential triggering kernel. Here, the self-excitation parameter $\alpha_{s}$ 
and the time constant $\tau_{s}$ depend on the spatiotemporal pattern $s$.
We assume that $\alpha_{s}$ is sampled from ${\operatorname{Ga}}(\alpha_{\rm time}, \beta_{\rm time})$ and 
that $\tau_s$ is drawn from a uniform prior.


\subsection{Content Distribution}
The content associated to each geolocated post is represented by a vector $\mathbf{d}$ in which each element is a word sampled from a 
vocabulary $\Vcal$ as
\begin{equation}
d_j \sim \rm{Multinomial}(\thetab_s)
\end{equation}
where $\thetab_s$ is a $|\Vcal|$-length vector whose elements encode each word probability and which depends on the spatiotemporal pattern $s$ that the 
post belongs to.
For each spatiotemporal pattern we assume $\thetab_s$ is sampled from $\operatorname{Dirichlet}(\theta_0)$.


\subsection{Spatial Distribution}
The spatial location associated to each geolocated post is represented by a 2-dimensional vector $\mathbf{r} = (x,y)$, which is drawn from a
normal distribution, i.e.
\begin{equation}
\mathbf{r} \sim \Ncal(\mathbf{R}_s, \mathbf{\Sigma}_s)
\end{equation}
where the mean and covariance depend on the spatiotemporal pattern $s$ the post belongs to.
Here, we consider isotropic distributions with $\Sigmab_s = \sigma_s^2 \mathbf{I}$ and assume that for each spatiotemporal
pattern $s$ the mean $\mathbf{R}_s$ is drawn from a flat (improper\footnote{Note that while this would be problematic in the generative setting, it presents 
no problems in the inferential setting.}) prior and $\sigma_s^2$ is drawn from ${\operatorname{Inv-Ga}}(1, \beta_{\rm space}^{-1}) = {\operatorname{Inv-Exp}(\beta_{\rm space}^{-1})}$. 
The conjugacy of these priors will be crucial in designing an efficient inference procedure in Section~\ref{sec:inference}.

\subsection{Generative Process}
In the previous sections we assumed that the spatiotemporal pattern a post belongs to is given and then modeled the time, content, and location
of the post using conditional distributions whose parameters depend on the associated pattern. 
Next we introduce a generative process---inspired by the Chinese Restaurant Process---in which posts are assigned to spatiotemporal patterns as 
they are generated:
\begin{itemize}
\setlength\itemsep{-0.3em}
\item[1.] Initialize the number of patterns: $S \to 0$.

\item[2.] For $n=1, \ldots, N$: 
\vspace{-.2cm}
\begin{itemize}\setlength\itemsep{-0.8em}
\item[(a)] Sample $t_n$ from Poisson($\lambda_0+\sum_{i=1}^{n-1}\gamma_{s_i}(t_n,t_i)$) \\
\item[(b)] 
\begin{itemize}
\item[(i)]
Assign post $n$ to a new spatiotemporal pattern $s_n = S+1$ with probability 
$$\frac{\lambda_0}{\lambda_0+\sum_{i=1}^{n-1}\gamma_{s_i}(t_n,t_i)},$$ 
draw its associated parameters $(\thetab_{s_n}, \mathbf{R}_{s_n}, \sigma_{s_n}^{-2})$ from the 
corresponding priors, and increase the number of patterns $S \to S+1$.

\item[(ii)]
If post $n$ is not assigned to a new pattern in step (i) above, then with probability 
$$\frac{\lambda_{k}(t_n)}{\lambda_0+\sum_{i=1}^{n-1}\gamma_{s_i}(t_n,t_i)},$$ 
assign it to an existing spatiotemporal pattern $s_n = k$, where $k = 1, \ldots, S$ and
 $\lambda_{k}(t_n) = \sum_{i=1}^{n-1}\gamma_{s_i}(t_n,t_i) \II(s_i = k)$ and reuse $(\thetab_k, \mathbf{R}_k, \sigma_k)$. \\
\end{itemize}

\item[(c)] Sample $\mathbf{d}_n$ from $\rm{Multinomial}(\thetab_{s_n})$ and $\rb_n$ from  $\Ncal(\mathbf{R}_{s_n}, \sigma_{s_n}^2\mathbf{I})$.
\end{itemize}
\end{itemize}
Note that the base intensity $\lambda_0$ controls the rate at which new patterns are created.

\begin{figure*}[!t]
\centering
\subfloat[Accuracy of spatiotemporal pattern inference]{\includegraphics[width=0.46\textwidth]{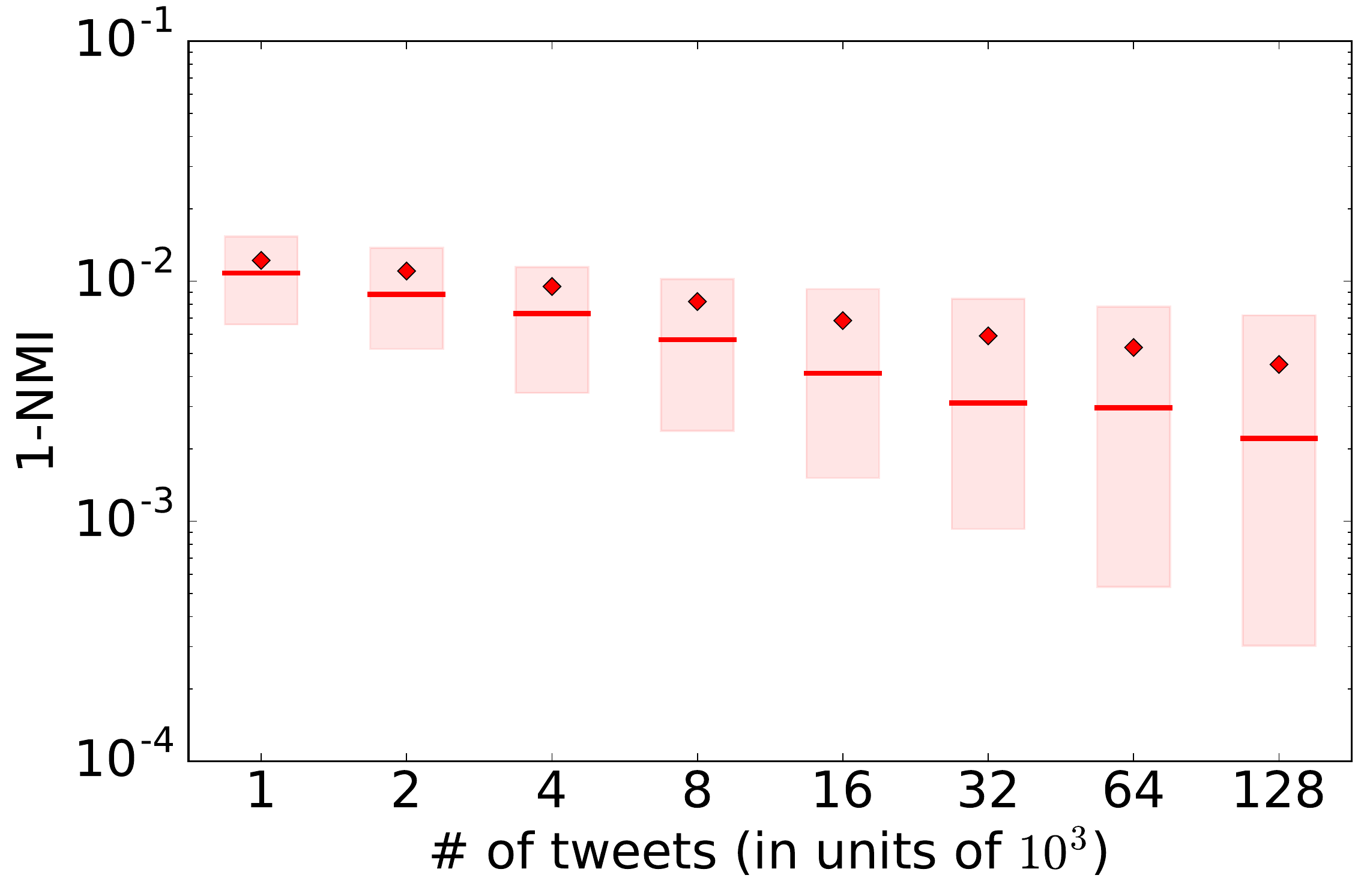} \label{fig:nmi}} \hspace{3mm}
\subfloat[Estimation of $\alpha_s$]{\includegraphics[width=0.46\textwidth]{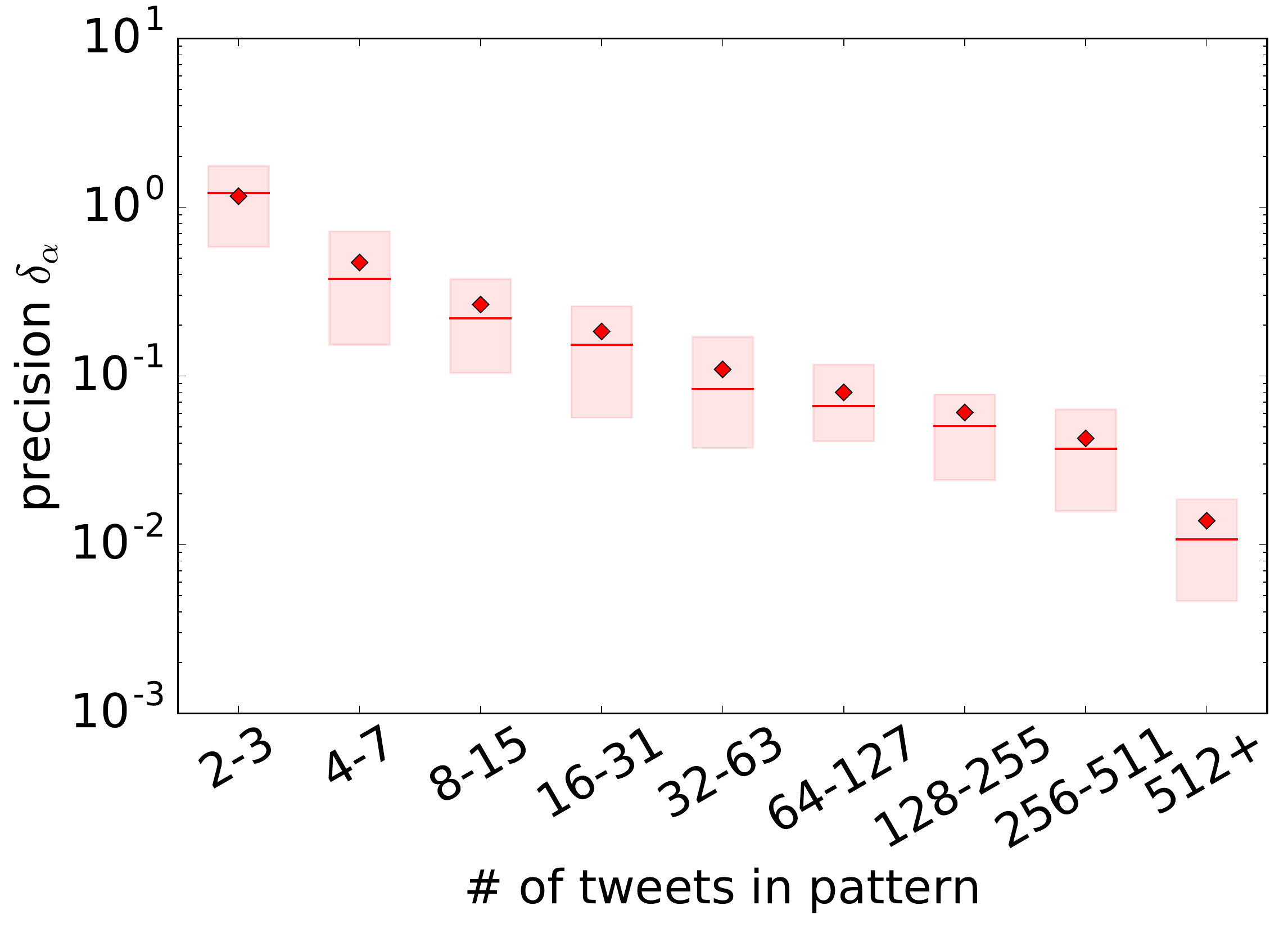} \label{fig:alpha_precision}}
\caption{Accuracy of our inference algorithm. Panel (a) shows the normalized mutual information (NMI) as a function of the number of 
tweets fed into the inference algorithm. Inferential accuracy remains high as the size of the dataset increases.
Panel (b) shows how the precision with which the self-excitation parameter $\alpha_s$ can be inferred increases as the number of observed tweets per inferred pattern increases. For both figures the diamonds mark the mean and the thick lines mark the mode, while each box extends from the lower to upper quartile.} \label{fig:synthetic-inference1}
\end{figure*}

\begin{figure*}[!t]
\centering
\subfloat[NMI vs. $\sigma_0$]{\includegraphics[width=0.46\textwidth]{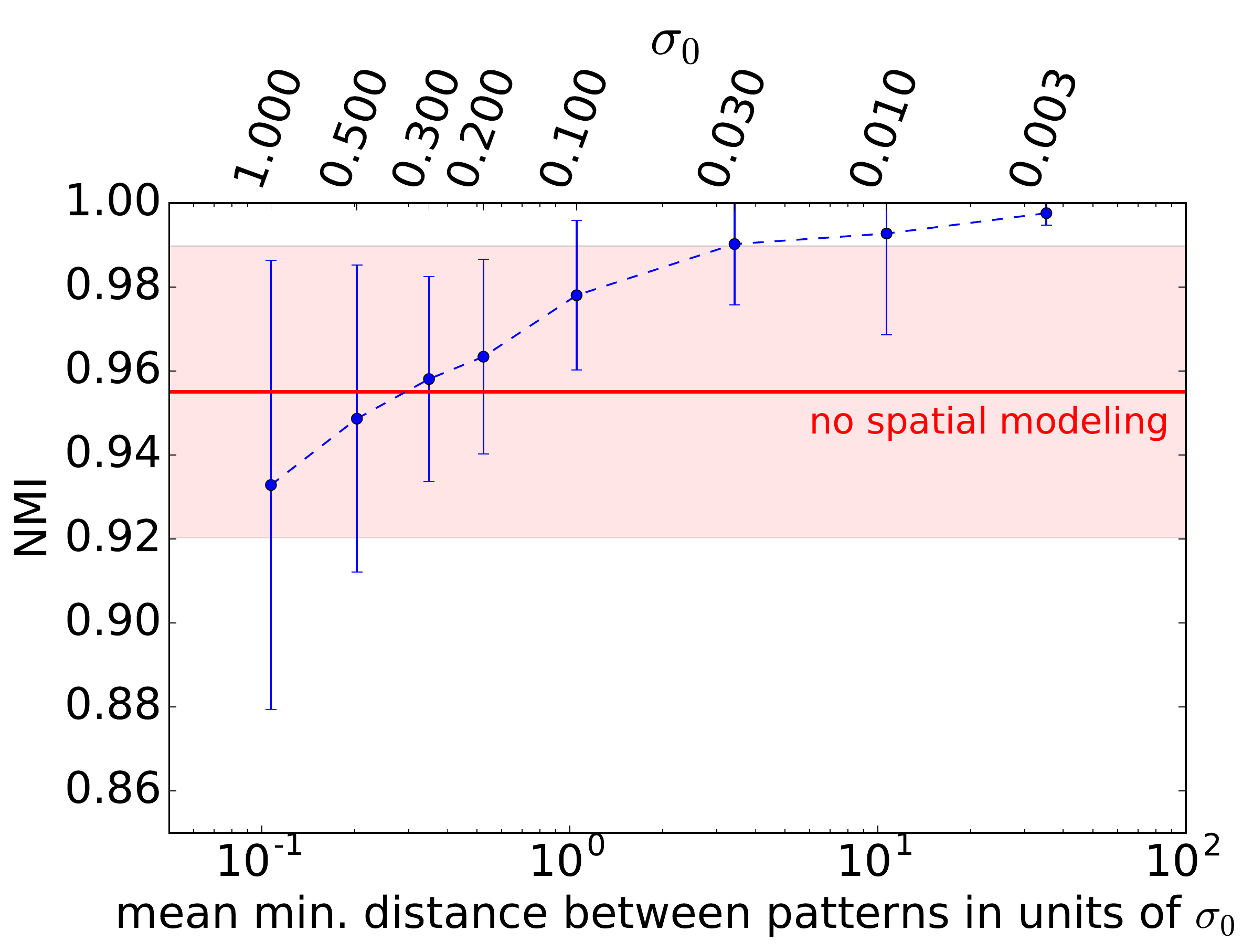} \label{fig:nmi_with_space}} \hspace{3mm}
\subfloat[SDHP vs. DHP]{\includegraphics[width=0.46\textwidth]{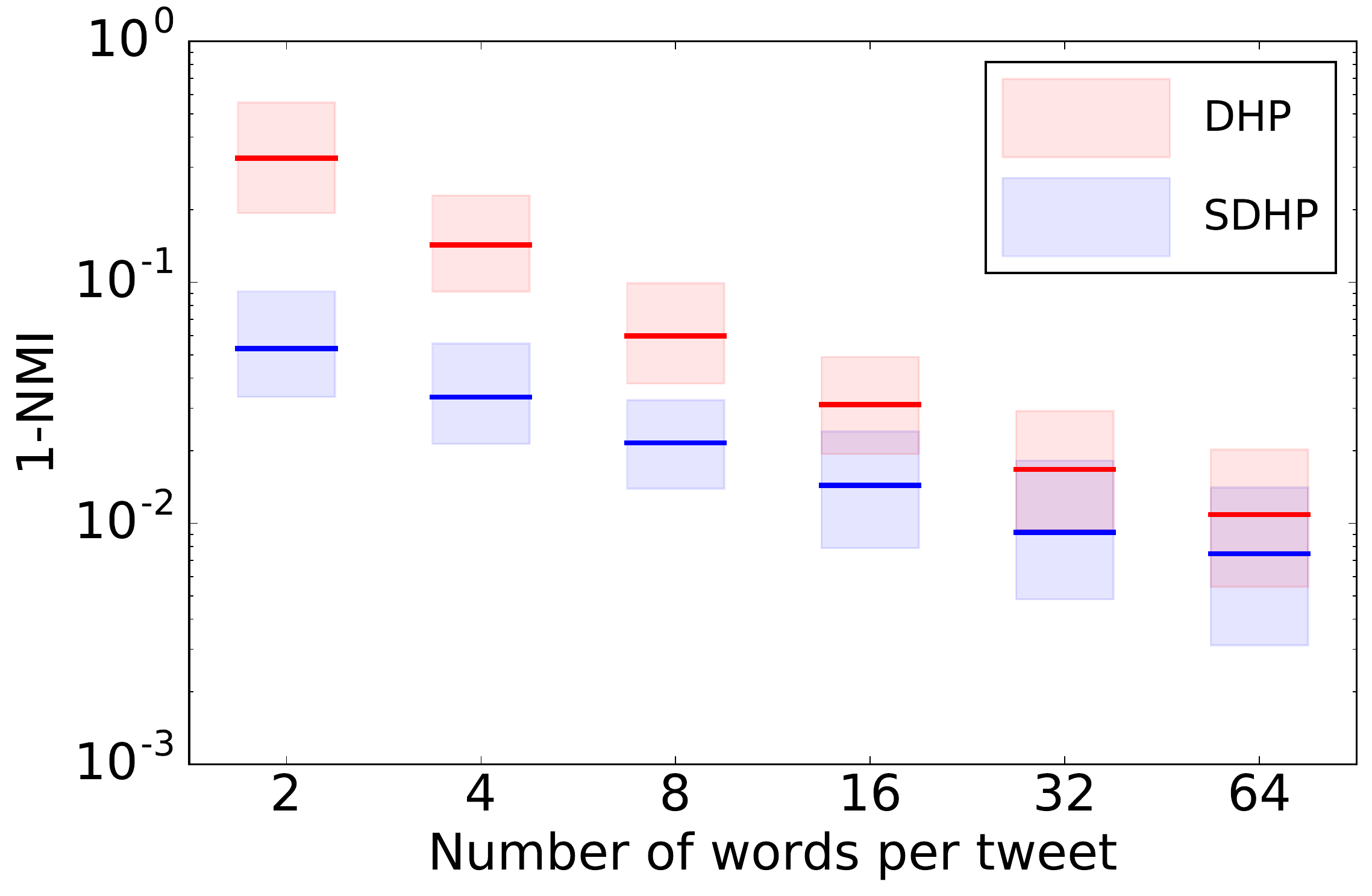} \label{fig:nwords_dep}}
\caption{To what extent does spatial information help during inference? Results for the SDHP are in blue, while results for the DHP are in red. 
Panel (a) shows the mean normalized mutual information (NMI) with standard errors
as a function of the true spatial variance $\sigma_0^2$ of the generated spatiotemporal patterns (top axis) as well as a function of a measure of spatial overlap (bottom axis).
Panel (b) shows $1-$NMI as a function of the number of words per tweet. Thick lines denote the mode, while each box stretches from the lower to upper quantile.} \label{fig:synthetic-inference2}
\end{figure*}

\section{Inference}
\label{sec:inference}

Given a collection of $N$ observed geolocated posts generated by the users of an online social network or media site during a time period $[0, T)$, 
our goal is to infer the spatiotemporal patterns that these posts belong to.
To efficiently sample from the posterior distribution, we derive a sequential Monte Carlo (SMC) algorithm \cite{gordon1993novel,doucet2009tutorial} that exploits the temporal dependencies in the observed
data to sequentially sample the latent variables associated to each geolocated post. The runtime of the algorithm is $\mathcal{O}(N \!\!\times\!\! |\mathcal{P}| \!\!\times\!\! \bar{S})$, where $|\mathcal{P}|$ is the number of particles
used during SMC and $\bar{S}$ is the mean number of latent spatiotemporal patterns.
For a detailed description of the algorithm please refer to the appendix.

\section{Experiments on Synthetic Data}
\label{sec:synthetic}
In this section we examine the performance of our inference algorithm under a variety of conditions using synthetic data drawn from the generative process. 

\subsection{Experimental Setup}
Each of the synthetic experiments described below has the same structure. 
All parameters but one---call it $x$---are kept fixed in each experiment,
while for each value of $x$ a large number of trials is done, with each trial running the (S)DHP
inference algorithm on a sample drawn from the generative process. Reported are summary statistics
for each set of trials corresponding to each value of $x$.\footnote{See appendix for more details on experimental setup.}

\subsection{Results}
We first evaluate the performance of our inference algorithm as the number of geolocated tweets\footnote{
Here and throughout the rest of the paper we will refer to the posts that make up the dataset under consideration as tweets.} 
in each sample as well as the number of tweets per inferred pattern increases.
Figure~\ref{fig:synthetic-inference1} summarizes the results, 
which show that: (i) the assignment of tweets to spatiotemporal patterns, 
measured by means of the normalized mutual information (NMI) between the true and inferred 
spatiotemporal patterns, remains accurate for large datasets; and (ii) the estimation of the self-excitation parameters $\alpha_s$, as measured by the precision
\begin{equation}
\delta_{\alpha} = \tfrac{| \alpha_s - \hat{\alpha}_s | }{ | \alpha_s + \hat{\alpha}_s | /2} \nonumber
\end{equation}
becomes more accurate as the size of inferred patterns increases, as expected. 

Next we evaluate the extent to which our inference algorithm benefits from the spatial information contained in each geolocated tweet. Intuitively, 
the smaller the spatial variance of a given pattern $s$, the more informative the spatial information of the tweets belonging to $s$
should be.
Figure~\ref{fig:nmi_with_space} confirms this intuition by showing the NMI as a function of the spatial variance $\sigma_0^2$ of the generated patterns.  As $\sigma_0$ decreases (so that the spatial overlap between different patterns decreases) the
inferential accuracy of the SDHP increases and surpasses that of the DHP~\cite{du2015dirichlet}. 
Moreover, Figure~\ref{fig:nwords_dep} shows that spatial information becomes more valuable when the associated content is less informative by comparing the 
NMI achieved by SDHP (in blue) and by the DHP (in red) for a varying number of words per tweet.

\section{Experiments on Real Data}
\label{sec:experiments-real}
In this section, we apply our model to real world data from Twitter. First, we show that our model recovers meaningful
spatiotemporal patterns across a diverse range of spatial and temporal scales. Then we demonstrate that the SDHP can be used
to perform accurate prediction of tweet locations, performing favorably with respect to two baselines~\cite{du2015dirichlet,heller2005bayesian}.
Finally we perform a quantitative evaluation of spatial and content goodness of fit in comparison to two baselines. 

\subsection{Experimental Setup}
Our data consist of the time, location and content of $12{,}558{,}046$ geolocated tweets posted during 
a one year period from January 1, 2013 to December 31, 2013 in a bounding box spanning Manhattan, 
New York City.
In this work, we focus on tweets written in English, as determined by the language tags provided by 
GNIP\footnote{\scriptsize \url{https://gnip.com/}}, and create two datasets: 
(i) $\Dcal_1$, which includes $|\Dcal_1| = 3{,}695{,}301$ tweets posted from Manhattan south of Central
Park; and
(ii) $\Dcal_2$, which includes $|\Dcal_2| = 69{,}538$ tweets posted from Central Park.
Finally, we preprocess the tweet content as follows: 
(i) all words are made lowercase; 
(ii) the $200$ most frequent words in the entire sample are filtered out; 
and (iii) punctuation/hashtags are left as is.
%

\subsection{Spatiotemporal patterns}
In this section, we perform a qualitative analysis of the spatiotemporal patterns picked out by the SDHP. To this end we
run our inference algorithm
on the Central Park dataset $\Dcal_2$, which reveals $7{,}639$ spatiotemporal patterns. Here we choose 
the hyperparameters by cross-validation on a held-out set and use the set of time constants
\mathchardef\mhyphen="2D
$\Psi_\tau = \{ {\rm hour}, {\rm day}, {\rm week}, {\rm month}, {\rm quarter \mhyphen year}, {\rm year}  \}$.

\begin{figure}
\centering
\includegraphics[width=0.9\textwidth]{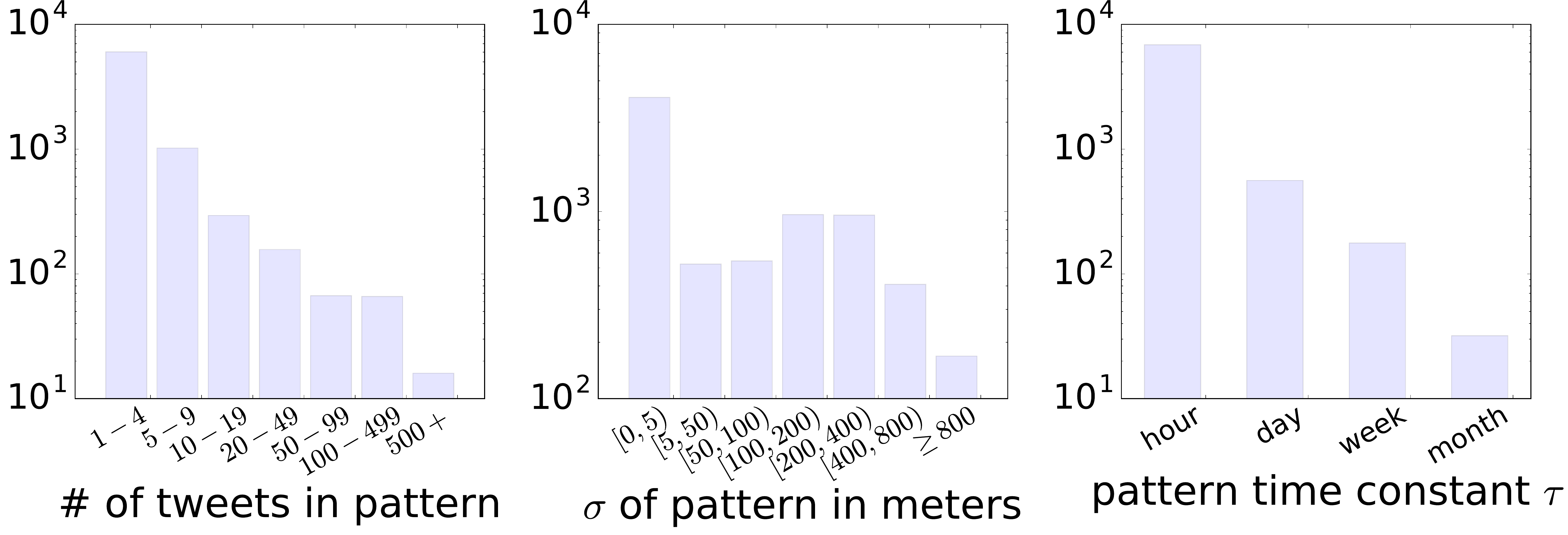}
\caption{The distribution of pattern sizes inferred across the entire dataset $\mathcal{D}_2$ in terms of number of tweets (left) and spatial extent in meters (middle). Also plotted (right) is the distribution of the time constants $\Psi_\tau$; note that none of the inferred patterns is assigned $\tau=$ quarter-year or $\tau=$ year.}
\label{fig:agg_topic_distros}
\end{figure}

First we examine the spatiotemporal characteristics of the revealed spatiotemporal patterns at an aggregate level. Intuitively, 
since Central Park encompasses a relatively large area where a variety of collective social activities are undertaken (e.g.~concerts, 
picnicking, sports, etc.), we expect to find patterns spanning a diverse range of spatial and temporal scales.
Figure~\ref{fig:agg_topic_distros} summarizes the results, which shows that the SDHP picks out spatiotemporal patterns with a diverse
range of pattern sizes (from a few tweets to hundreds), spatial extent (from $\sigma < 5$ to $\sigma > 1000$ meters), 
and time constants (from $\tau =$ hour to $\tau = $ month).

\begin{figure}[!t]
\centering
\includegraphics[width=0.7\textwidth]{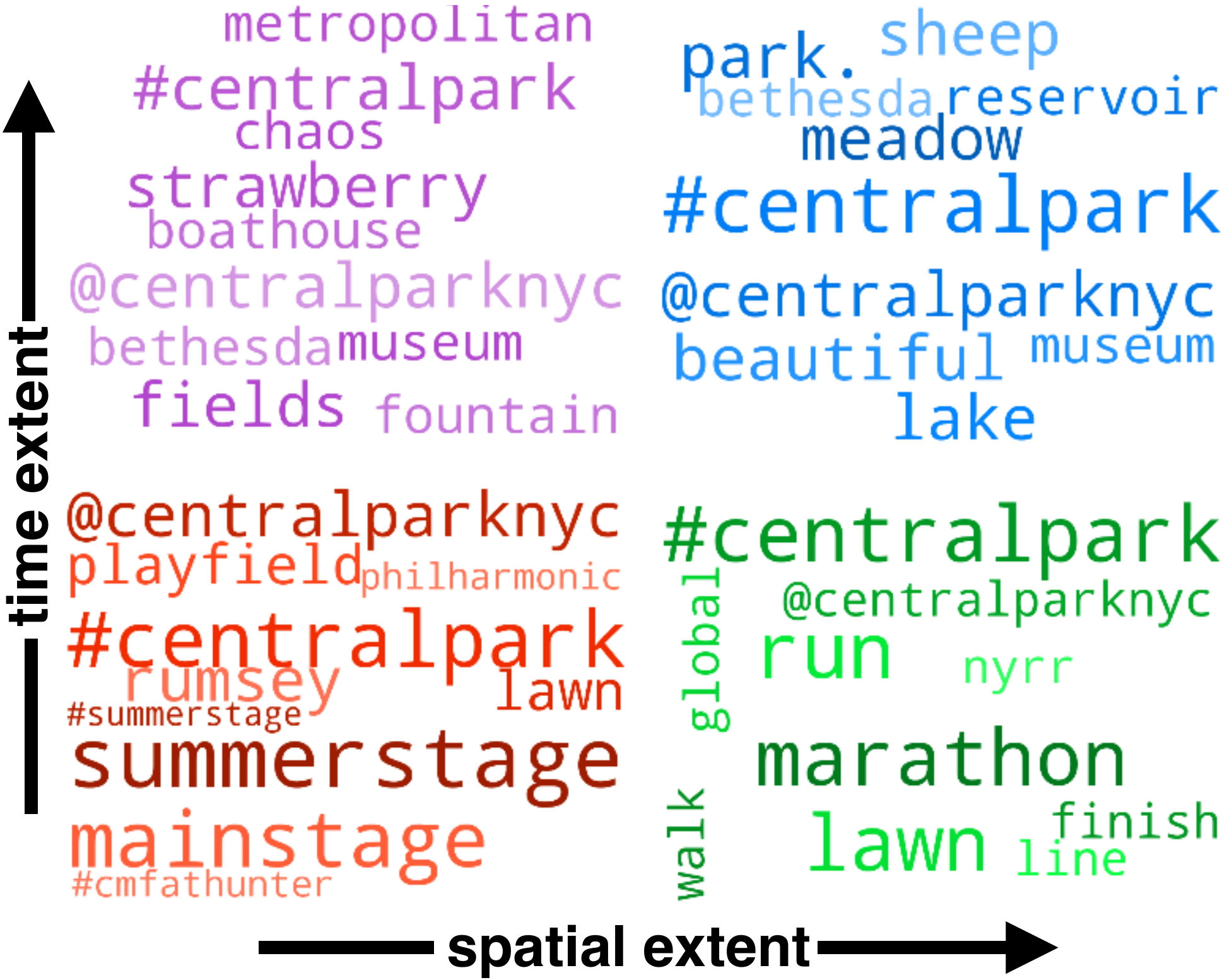}
\caption{Four word clouds showing how the content of inferred patterns varies according to their time duration (less than or greater than 1 week) and their spatial extent (less than or greater than 100 meters).}
\label{fig:aggwordcloud}
\end{figure}

\begin{figure*}[!t]
\centering
\subfloat[Content]{\includegraphics[width=0.60\textwidth]{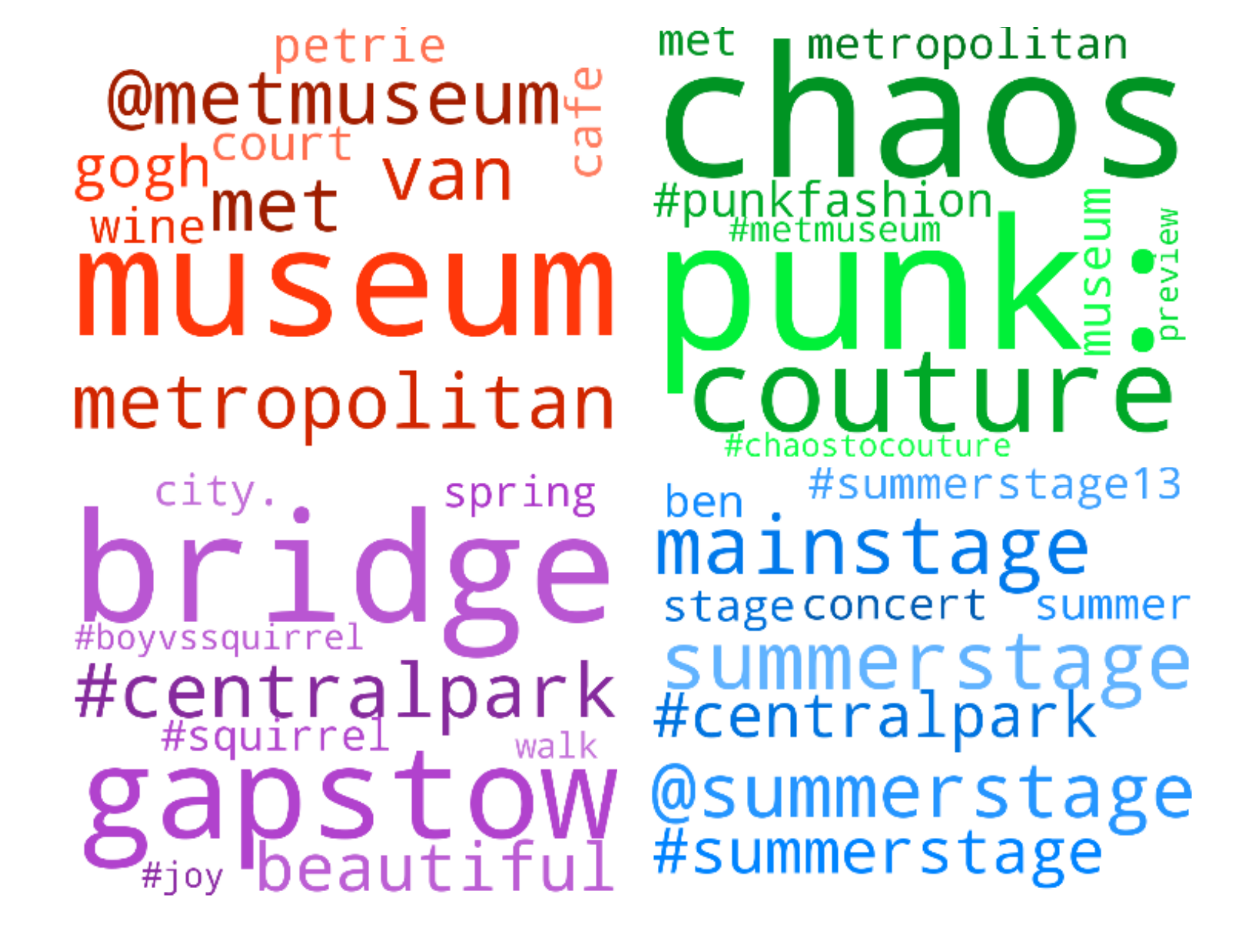}} \hspace{4mm}
\subfloat[Temporal Dynamics]{\includegraphics[width=0.72\textwidth]{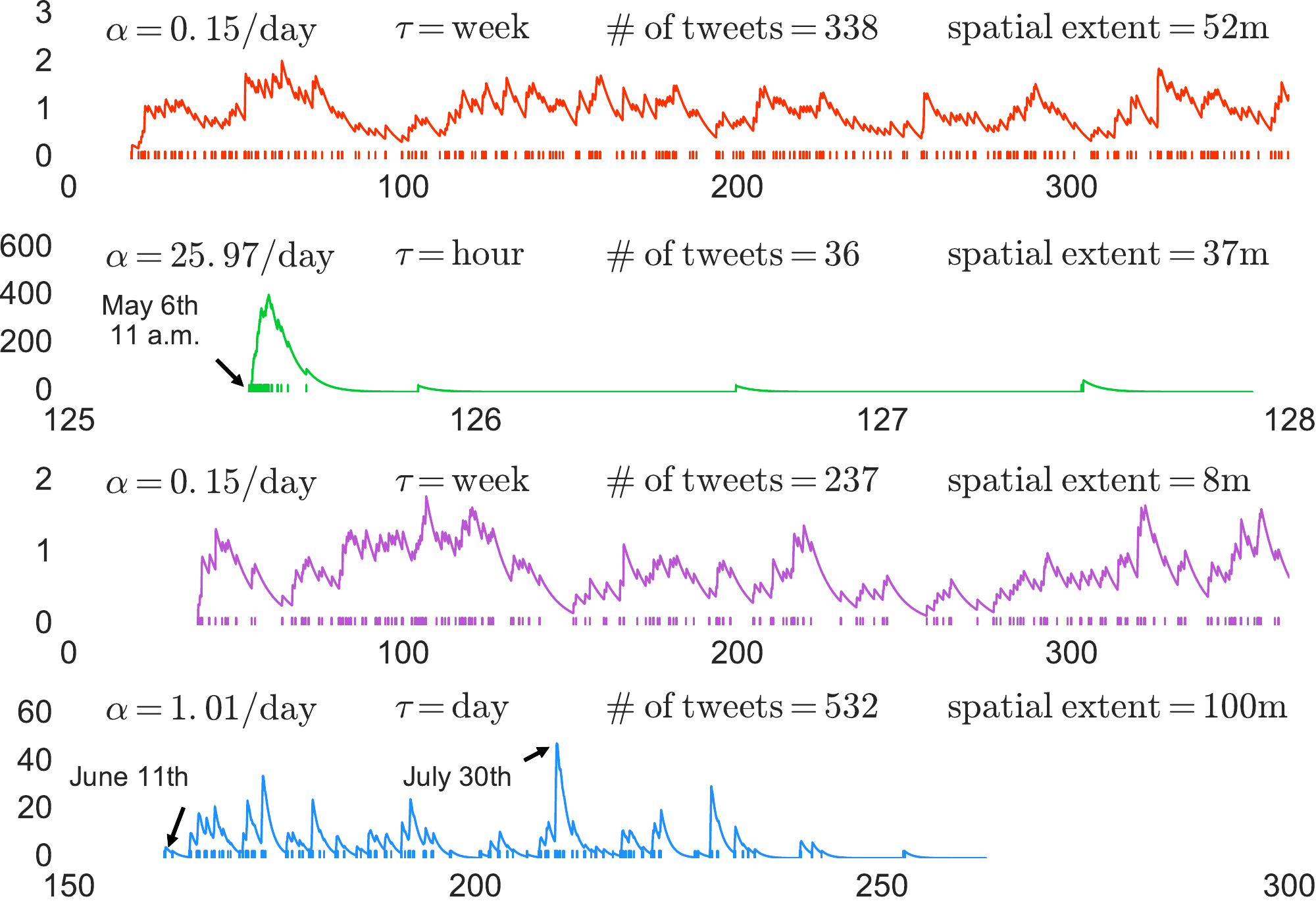}}
\caption{Content and temporal dynamics of the four spatiotemporal patterns depicted in Figure~\ref{fig:central_park}.
Panel (a) shows the word content of each pattern. Panel (b) shows the temporal dynamics for each pattern by means of the fitted intensities 
$\lambda_s(t)$ in units of inverse days and includes additional pattern-level summary statistics. Blips along the time axis correspond to individual tweets. 
The time axis is in units of days, while the pattern-specific intensities are in units of inverse days.
}
\label{fig:four_content_temporal_patterns}
\end{figure*}

Next we examine the content for patterns spanning different temporal and spatial scales. To do so we group the inferred spatiotemporal
patterns into four sets based on their spatial extent ($\sigma \lessgtr 100$m) and their time duration ($\Delta t \lessgtr 1$ week). 
Then for each set we create a word cloud that encodes the ten most frequent terms in the set, see Figure~\ref{fig:aggwordcloud}. 
Remarkably, the most popular words in each set reflect the underlying spatiotemporal characteristics. For example,
the spatiotemporal patterns that are localized in time ($\Delta t < 1$ week) often 
correspond to specific events which take place at specific locations in the park (e.g.~summer stage for $\sigma < 100$m) as well
as throughout the entire park (e.g.~the NYC Marathon for $\sigma > 100$m);
spatiotemporal patterns that unfold over longer periods of time ($\Delta t \ge 1$ week) often include words describing \emph{static} features
of Central Park such as `strawberry' and `fields' for $\sigma < 100$m, referring to a small memorial dedicated to John Lennon, or `reservoir' and `lake' for $\sigma \ge 100$m.
Finally we turn to examine the spatiotemporal characteristics of four individual patterns more closely: a typical visit to the Metropolitan Museum of Art (the Met), 
a press preview to an exhibition at the Met,\footnote{\scriptsize \url{http://www.metmuseum.org/press/exhibitions/2012/punk-chaos-to-couture}} a typical visit to the iconic Gapstow Bridge, and musical concerts held throughout the summer at Rumsey Playfield and Naumburg Bandshell. 
Figure~\ref{fig:central_park} shows the location of the tweets assigned to each spatiotemporal pattern, and Figure~\ref{fig:four_content_temporal_patterns} shows the content 
and temporal dynamics of each pattern by means of word clouds and fitted temporal intensities.
In terms of spatial dynamics, the collective social activities span a wide range, from $\sigma \approx 8$m for visits to Gapstow Bridge to 
$\sigma \approx 100$m for the summer concerts.
For the temporal dynamics, we find similar diversity, from $\alpha = 25.97$ and $\Delta t \approx 2$ days for the press preview to $\alpha = 1.01$ and 
$\Delta t \approx 91$ days for the summer concerts.
%
%
\subsection{Location Prediction}
\label{sec:locprediction}
We evaluate the performance of our model at predicting the location of tweets using dataset $\Dcal_1$, which corresponds to
tweets in Manhattan south of Central Park.
We compare the performance of the SDHP in this context against two baselines: Bayesian Hierarchical Clustering (BHC\footnote{For the BHC
we consider isotropic normal spatial \emph{and} temporal distributions with a normal-gamma prior and dirichlet-multinomial distributions for the content.})
and the Dirichlet-Hawkes Process (DHP)~\cite{du2015dirichlet}. Note that although the SHDP was not specifically designed for the purpose of tweet location prediction---rather it was designed to discover spatiotemporal patterns more broadly---we expect that if the assumptions underlying the spatial modeling are appropriate to the given dataset, then the SDHP should be able to make reasonable location predictions.\footnote{At least for those tweets belonging to sufficiently large and well-localized patterns.}

In this section, our focus is on tweets that belong to spatiotemporal patterns with sufficiently tight spatial structure such that the locations of the corresponding tweets are
\emph{predictable}.
To this end, we will experiment with five subsets of length $2{,}500$ tweets,\footnote{We limit each subset to $2{,}500$ tweets because BHC does not scale to larger 
datasets.} built by filtering on a fixed lists of keywords: `gallery', `park', `concert', `sports', and `bar'. 
Note that despite the potential for tight spatial structure of the underlying patterns suggested by the chosen keywords, 
the tweets within each subset are spread throughout Manhattan, as shown
in Figure~\ref{fig:location_prediction}. 
For example, in the case of `gallery', which represents the most concentrated sample of tweets, while there is a large cluster of galleries in Chelsea, there are also many galleries in SoHo, the Lower East Side, Midtown, and elsewhere.
This, combined with the fact that an individual gallery may receive only a handful of---or even a single---mentions makes the prediction task difficult.
\newcommand{\mapfigwidth}{0.185}
\begin{figure*}[!t]
\centering
\subfloat[Gallery]{\includegraphics[width=\mapfigwidth\textwidth]{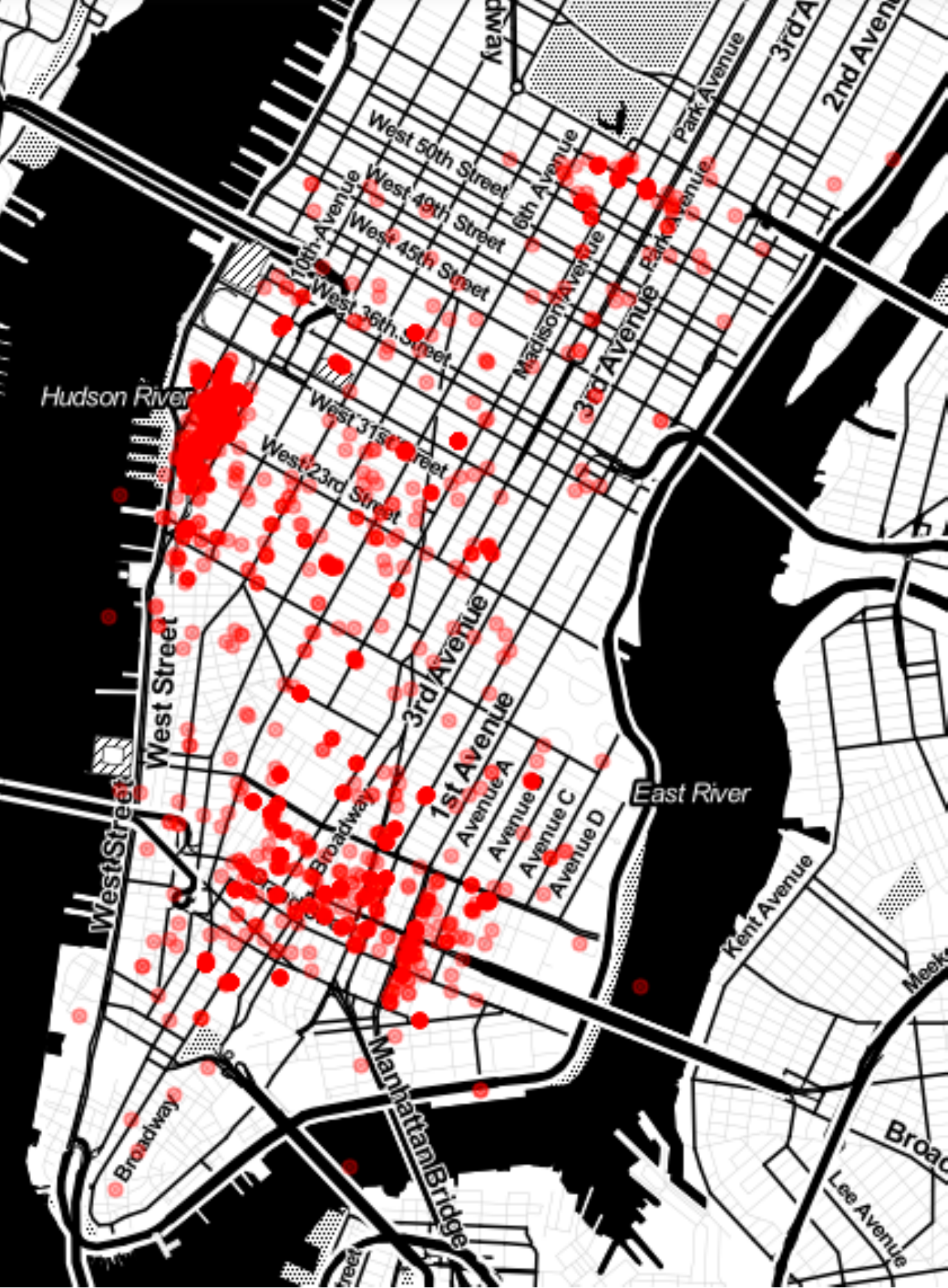}} \hspace{1mm}
\subfloat[Bar]{\includegraphics[width=\mapfigwidth\textwidth]{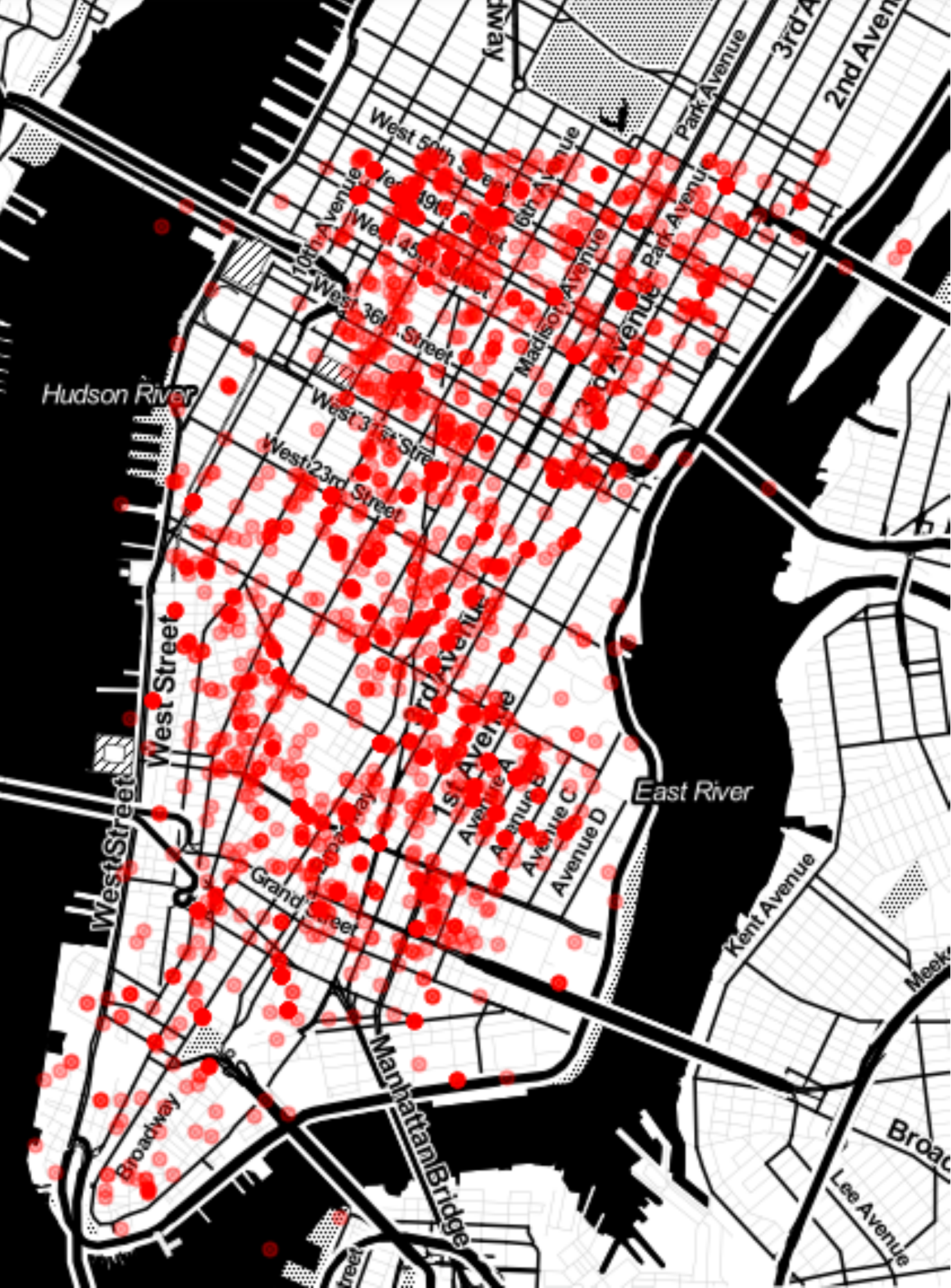}} \hspace{1mm}
\subfloat[Concert]{\includegraphics[width=\mapfigwidth\textwidth]{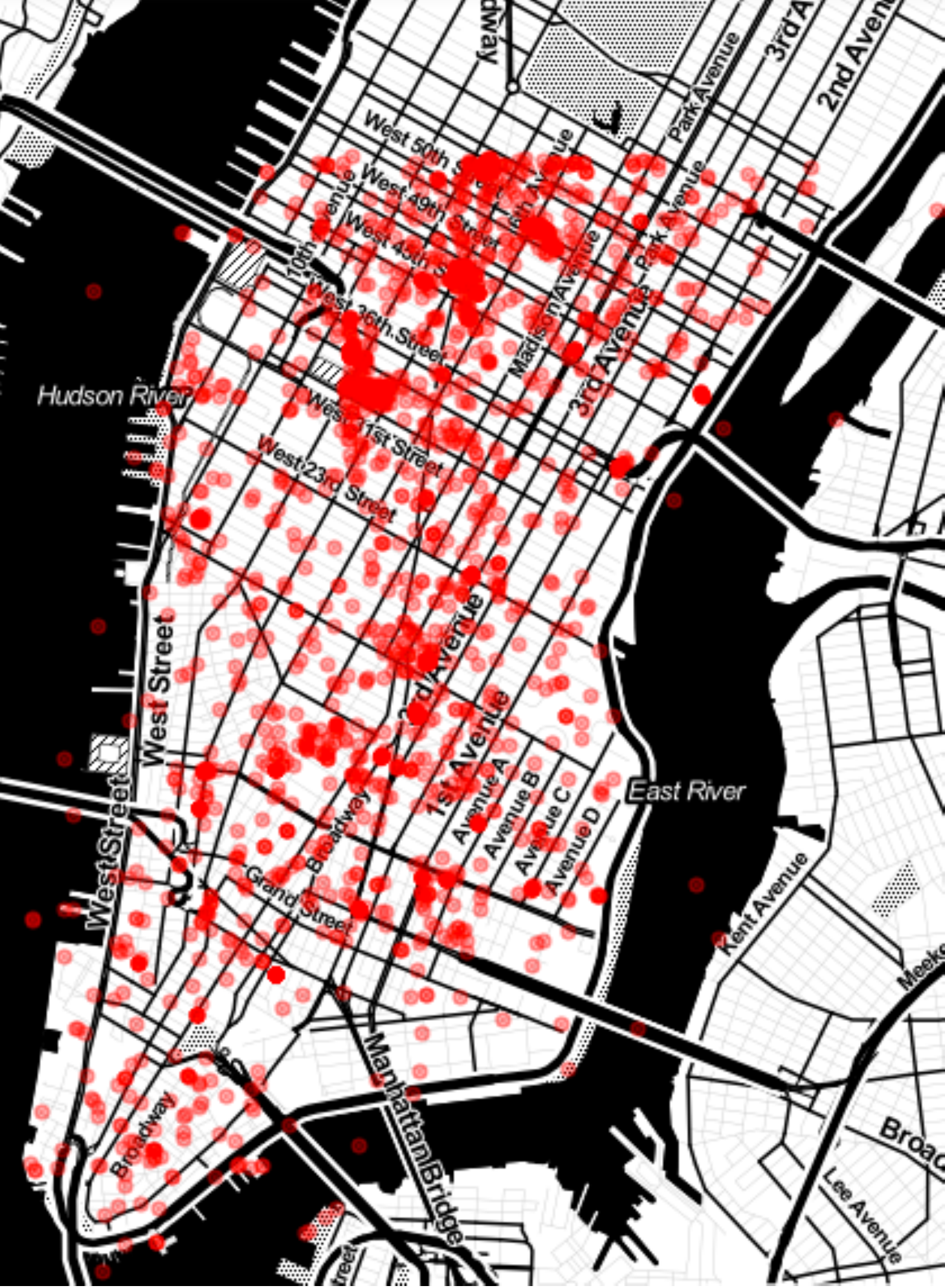}} \hspace{1mm}
\subfloat[Park]{\includegraphics[width=\mapfigwidth\textwidth]{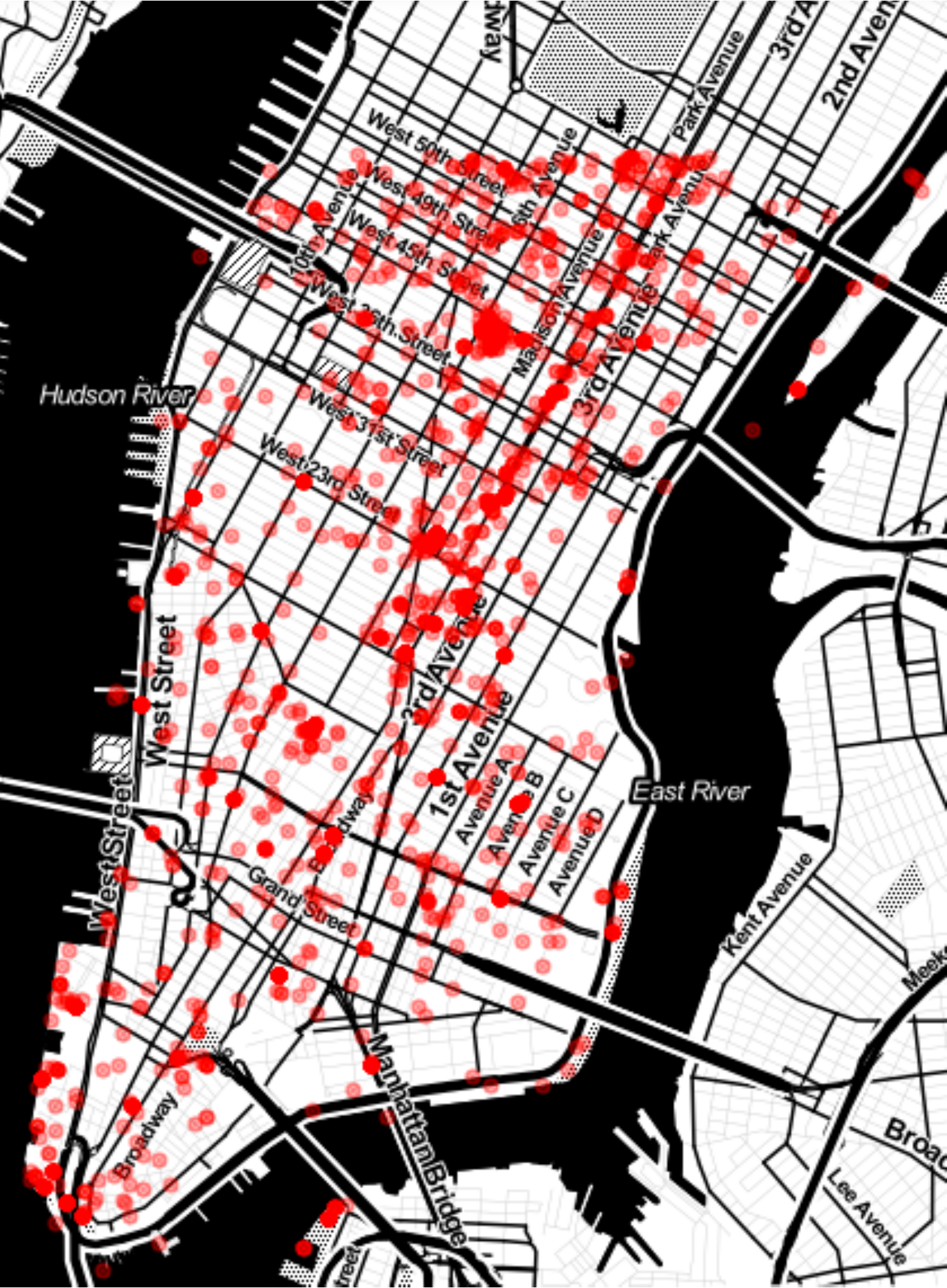}} \hspace{1mm}
\subfloat[Sports]{\includegraphics[width=\mapfigwidth\textwidth]{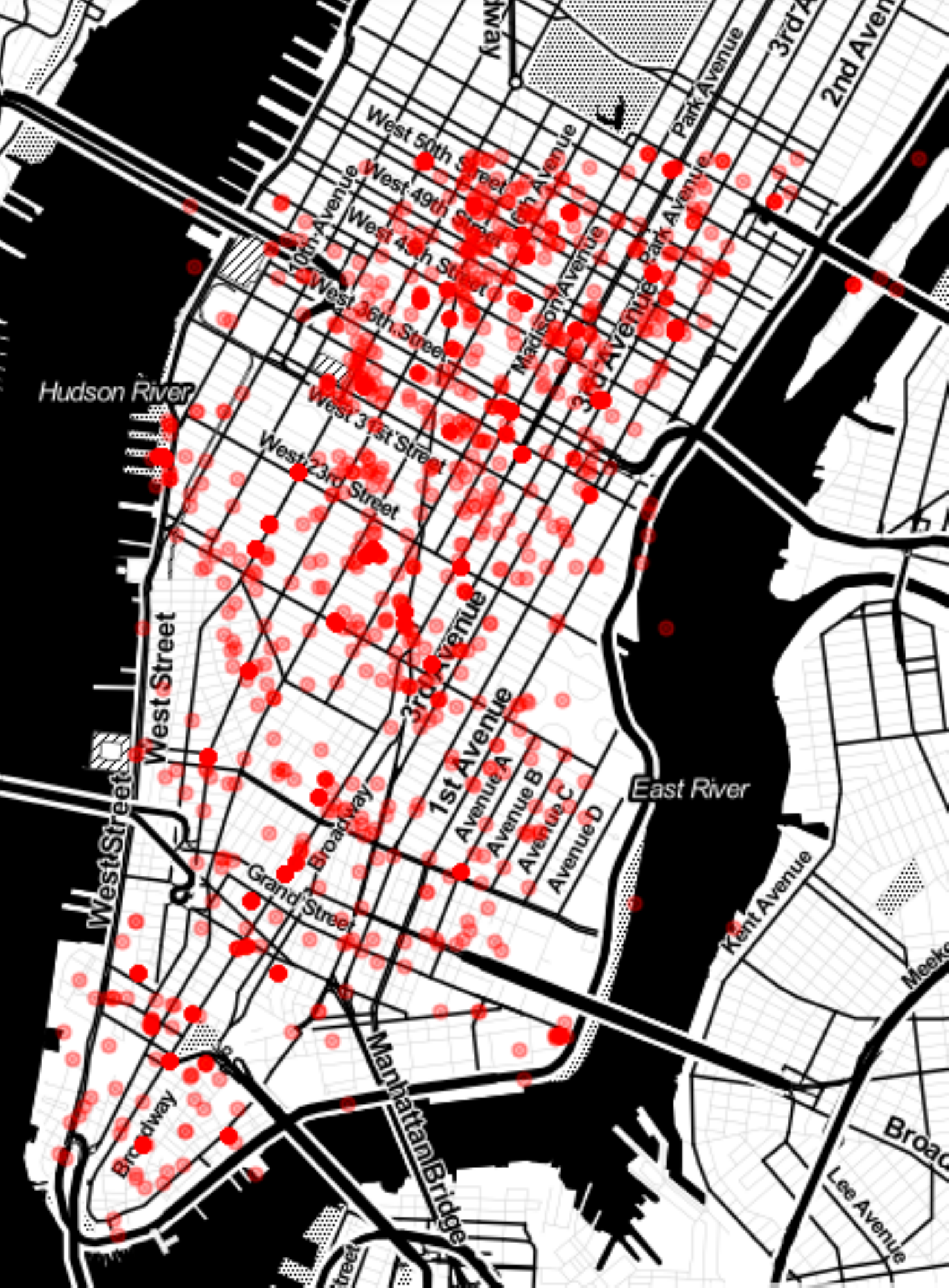}} \hspace{1mm}
\caption{Locations of the tweets in the five subsets of dataset $\Dcal_2$ defined by the keywords `gallery', `bar', `concert', `park', and `sports', respectively, that 
are used in the experiments in Sec.~\ref{sec:locprediction} and Sec.~\ref{sec:gof}.}
\label{fig:location_prediction}
\end{figure*}

For each subset, we proceed as follows:
\begin{itemize}\setlength\itemsep{0em}
\item[1.] We hide the location of 2\% of the tweets, picked at random, excluding the first (chronologically) 20\% of the tweets (burn-in period).
\item[2.] We run the corresponding inference algorithm.
\item[3.] We infer the position of each hidden tweet via its spatial posterior, which is just the mean location of the other (non-hidden)
tweets in the spatiotemporal pattern the tweet is assigned to.
\item[4.] We repeat steps $1$-$3$ for $100$ trials. If a tweet is hidden in more than one trial, we infer its position using the trial in which
it is assigned to the tightest pattern (lower $\sigma_s$).
\end{itemize}
In the above procedure, we set the hyperparameters of all methods using a held-out validation set built using the `gallery' subset and
set $\Psi_\tau = \{ {\rm hour}, {\rm day}, {\rm week}, {\rm month}, {\rm quarter \mhyphen year} \}$.

Then we evaluate the performance of each modeling framework using the root mean square error (RMSE), applying two different 
selection criteria that consider only location predictions about which the corresponding modeling framework is reasonably certain 
(see appendix for details).

Table~\ref{table:hide} summarizes the results by means of the RMSE in units of the square root of the spatial variance of the entire 
subset of tweets, which ranges from  $\sim\!\!1370$m in the case of `gallery' to $\sim\!\!1660$m in the case of `park'. For most of the samples
and both of the selection criteria, SDHP gives better location predictions than the two baselines. 
The accuracy of the prediction, however, varies significantly from
sample to sample, as the difficulty of the task varies substantially depending on the keyword. Note that in the case of `concert' too few tweets pass the tight selection criteria for the measure to be computed. 
\newcommand{\DHPGallLoose}{0.087}
\newcommand{\DHPGallTight}{0.047}
\newcommand{\DHPParkLoose}{0.055}
\newcommand{\DHPParkTight}{0.057}
\newcommand{\DHPBarLoose}{\bf 0.021}
\newcommand{\DHPBarTight}{0.239}
\newcommand{\DHPSporLoose}{0.388}
\newcommand{\DHPSporTight}{0.600}
\newcommand{\DHPConcLoose}{0.088}
\newcommand{\DHPConcTight}{---}
\newcommand{\SDHPGallLoose}{\bf 0.056}
\newcommand{\SDHPGallTight}{0.167}
\newcommand{\SDHPParkLoose}{\bf 0.051}
\newcommand{\SDHPParkTight}{\bf 0.040}
\newcommand{\SDHPBarLoose}{0.063}
\newcommand{\SDHPBarTight}{\bf 0.029}
\newcommand{\SDHPSporLoose}{\bf 0.247}
\newcommand{\SDHPSporTight}{\bf 0.289}
\newcommand{\SDHPConcLoose}{\bf 0.080}
\newcommand{\SDHPConcTight}{---}
\newcommand{\BHCGallLoose}{0.068}
\newcommand{\BHCGallTight}{\bf 0.013}
\newcommand{\BHCParkLoose}{0.111}
\newcommand{\BHCParkTight}{0.134}
\newcommand{\BHCBarLoose}{0.134}
\newcommand{\BHCBarTight}{0.145}
\newcommand{\BHCSporLoose}{0.429}
\newcommand{\BHCSporTight}{0.449}
\newcommand{\BHCConcLoose}{0.132}
\newcommand{\BHCConcTight}{---}

%
\vspace{-.2cm}
\subsection{Goodness of fit}
\label{sec:gof}
\begin{table}[t]
\definecolor{mygray}{rgb}{0.8,.8,.8}
\resizebox{.9\columnwidth}{!}{%
\begin{tabular}{|*{7}{c|}}  
\hline
 \cellcolor{mygray} & \multicolumn{3}{c }{\bf loose selection} & \multicolumn{3}{|c|}{\bf tight selection} \\ \hline 
\multicolumn{1}{|c|}{{\bf keyword}} & SDHP & DHP & BHC & SDHP & DHP & BHC \\ \hline
 gallery        & \SDHPGallLoose          &  \DHPGallLoose        &  \BHCGallLoose   & \SDHPGallTight      & \DHPGallTight        &  \BHCGallTight \\ 
 park            &  \SDHPParkLoose        & \DHPParkLoose       & \BHCParkLoose   &  \SDHPParkTight     & \DHPParkTight       & \BHCParkTight \\ 
 concert       & \SDHPConcLoose        & \DHPConcLoose       & \BHCConcLoose  & \SDHPConcTight     & \DHPConcTight      & \BHCConcTight \\ 
 sports         &  \SDHPSporLoose       & \DHPSporLoose       & \BHCSporLoose    &  \SDHPSporTight     &  \DHPSporTight      & \BHCSporTight \\ 
 bar              &  \SDHPBarLoose        & \DHPBarLoose         & \BHCBarLoose       &  \SDHPBarTight      & \DHPBarTight         & \BHCBarTight \\  \hline 
 \end{tabular}
 }  
\caption{\label{table:hide}Results for tweet location prediction.}
%
\vspace{.2cm}
%
\definecolor{mygray}{rgb}{0.8,.8,.8}
\resizebox{.9\columnwidth}{!}{%
\begin{tabular}{|*{5}{c|}}  
\hline
 \cellcolor{mygray} & \multicolumn{2}{c }{\begin{tabular}{@{}c@{}}{\bf spatial goodness of fit} \\ mean log probability\end{tabular}} & \multicolumn{2}{|c| }{\begin{tabular}{@{}c@{}}{\bf content goodness of fit} \\ perplexity per word \end{tabular}} \\ \hline 
\multicolumn{1}{|c|}{{\bf keyword}} & SDHP & GMM & SDHP & DHP \\ \hline
 gallery       & 6.26         &  \bf 9.15          & $\mathbf{4.86\times 10^4}$       & 5.36$\times 10^4$                   \\          
 park           & 5.64         &  \bf 8.84          & $\mathbf{12.1\times 10^4}$       & $16.0\times 10^4$                   \\          
 concert      & 2.96         &  \bf 6.77          & 18.6$\times 10^4$                      &  $\mathbf{18.0 \times 10^4}$   \\        
 sports        & 3.88         &  \bf 8.77          & $\mathbf{10.3 \times 10^4}$       & 10.9$\times 10^4$                   \\        
 bar             & 4.17         &  \bf 6.25          & $\mathbf{16.3 \times 10^4}$      & 18.3$\times 10^4$                    \\ \hline
\end{tabular} 
}
\caption{\label{table:gof}Results for goodness of fit tests.}
\end{table}
We evaluate the goodness of fit of the SDHP at modeling collective social activity both in terms of spatial and content dynamics using the five
subsets of dataset $\Dcal_1$ described in Sec.~\ref{sec:locprediction}. 
In terms of spatial dynamics, we compare the SDHP to a Gaussian Mixture Model (GMM), 
while in terms of content dynamics, we compare the SDHP to the Dirichlet Hawkes Process (DHP).
In both cases the goodness of fit measure is a form of mean marginal log likelihood (see the appendix for details).
Table~\ref{table:gof} summarizes the results for both goodness of fit tests.

We see that for all five keywords the SDHP lags behind the GMM in the goodness of the spatial fit, but that the SDHP still provides a reasonable fit. 
This is as expected,
since the SDHP also pays close attention to temporal and content dynamics; e.g.~as we saw in Figure~\ref{fig:central_park} in the case of the two patterns located 
at the Met, the SDHP will tend to model a given spatial cluster with multiple patterns if they have varying temporal/content characteristics, while the GMM will prefer a sparser representation and use the additional gaussian component(s) to improve the fit elsewhere. 

For the content goodness of fit we see that the SDHP outperforms the DHP in four out of five samples (underperforming only for the keyword with the lowest
spatial goodness of fit, namely `concert'). This is encouraging, as it demonstrates that utilizing spatial information allows the SDHP to a provide a more
faithful content model for real world data.

\section{Conclusions}
\label{sec:conclusions}
We proposed a novel continuous spatiotemporal probabilistic model for clustering streaming geolocated data,
the Spatial Dirichlet Hawkes Process (SDHP), and developed an efficient, online inference algorithm based on sequential Monte Carlo
that scales to millions of posts. 
We showcased the efficacy of the model on data from Twitter and demonstrated that it can reveal a wide range of spatiotemporal
patterns underlying different collective social activities. In this application domain our model provides a better fit to the content dynamics 
of the data as well as more accurate location predictions than several alternatives~\cite{heller2005bayesian,du2015dirichlet}.

Our work opens up many interesting avenues for future work.
For example, we consider isotropic distributions to model the spatial dynamics of each pattern; however, there are many
scenarios (e.g.~traffic jams) in which other distributions may be more appropriate. 
It would also be valuable to augment our model to have a hierarchical design in which there are canonical spatiotemporal patterns 
(e.g.~traffic jams) with different spatial distributions that are then instantiated at different points in time and space.
In addition it would be interesting to apply our model to other sources of geolocated data, such as cellular phone data, mobile messaging 
data (e.g.~Whatsapp or Snapchat), or mobile photo-sharing data (e.g.~Instagram).

More broadly it could be useful to integrate our model into a larger information extraction pipeline, especially in the context of extracting
insights from urban social media data for the use of local government. For example, one could design a system that assists local prosecutors
in identifying possible cases of consumer fraud and tenant harassment.
Finally, whereas we have focused on collective social activities, it would be instructive to use the SDHP in the context of 
spatiotemporally localized \emph{events}, especially in the context of crisis and disaster awareness \cite{imran2013extracting}. 

 \section*{Acknowledgment}
The authors would like to thank Isabel Valera and Charalampos Mavroforakis for helpful discussions.

\bibliographystyle{abbrv}
\bibliographystyle{unsrt}
\small
\bibliography{bibi}

\section{Appendix}

\subsection{Details on Inference Algorithm}
In this subsection we describe how Sequential Monte Carlo can be used to infer latent patterns from the observed spatiotemporal and content data.
The posterior distribution $p(s_{1:n} | t_{1:n}, \mathbf{d}_{1:n}, \mathbf{r}_{1:n})$ is 
sequentially approximated from $n=1$ to $n=N$ with a set of $|\Pcal |$ particles that are sampled from a proposal distribution that factorizes as
\begin{align}
\nonumber
q_n(s_{\le n} | t_{\le n}, \mathbf{d}_{\le n}, \mathbf{r}_{\le n} ) &= q_n(s_n |  s_{<n},  t_{\le n}, \mathbf{d}_{\le n}, \mathbf{r}_{\le n}) \nonumber \\
& \times q_{n-1}(s_{< n} | t_{< n}, \mathbf{d}_{<n}, \mathbf{r}_{<n} ) \nonumber
\end{align}
where $q_n(s_n |  s_{<n},  t_{\le n}, \mathbf{d}_{\le n}, \mathbf{r}_{\le n})$ is given by
\begin{equation}
\label{eqn:proposal_distribution}
\frac{p(s_n | s_{<n}, t_{\le n}) p(\mathbf{d}_n | s_{\le n}, \mathbf{d}_{<n}) p(\mathbf{r}_n | s_{\le n}, \mathbf{r}_{<n})}{\sum_{s_n} p(s_n | s_{<n}, t_{\le n}) p(\mathbf{d}_n | s_{\le n}, \mathbf{d}_{<n}) p(\mathbf{r}_n | s_{\le n}, \mathbf{r}_{<n})} 
\end{equation}
In the above expression, the distribution $p(s_n | s_{<n}, t_{\le n})$ is given by  
 \begin{equation}
\label{eqn:topicprob}
p(s_n | s_{<n}, t_{\le n})  = \frac{\lambda_{s_n}(t_n)}{\lambda_0+\sum_{i=1}^{n-1}\gamma_{s_i}(t_n,t_i)}
\end{equation}
where the numerator $\lambda_{s_n}(t_n)$ is equal to $\lambda_0$ when $s_n$ is a new spatiotemporal pattern.

We can exploit the conjugacy between the multinomial and the Dirichlet distributions as well as the conjugacy between the normal distribution and normal-gamma prior to integrate out the word distributions $\thetab_s$ and spatial parameters $\{ \sigma_s, \mathbf{R}_s \}$, respectively, and obtain the marginal likelihoods:
\begin{equation}
p(\mathbf{d}_n | s_{\le n}, \mathbf{d}_{<n}) = 
\frac{\Gamma({C^{s_n /\ \!\!\mathbf{d}_n}} \!+\! V\theta_0) \prod_v^V \! \Gamma({C^{s_n /\ \!\!\mathbf{d}_n}_v} \!+\! {C^{\mathbf{d}_n}_v} \!+\! \theta_0)}
{\Gamma({C^{s_n /\ \!\!\mathbf{d}_n}} \!+\! {C^{\mathbf{d}_n}} \!+\! V\theta_0) \prod_v^V \! \Gamma({C^{s_n /\ \!\!\mathbf{d}_n}_v} \!+\! \theta_0)} 
 \nonumber
\end{equation}
where $V$ is the size of the observed vocabulary, ${C^{s_n /\ \!\!\mathbf{d}_n}}$ is the total number of words in the spatiotemporal pattern $s_n$ seen so far 
excluding $\mathbf{d}_n$, ${C^{s_n /\ \!\!\mathbf{d}_n}_v}$ is the total count for word $v$ in spatiotemporal pattern $s_n$ so far excluding $\mathbf{d}_n$, $C^{\mathbf{d}_n}_v$ 
is the total count for word $v$ in $\mathbf{d}_n$ and $C^{\mathbf{d}_n}$ is the total word count for $\mathbf{d}_n$; and,
\begin{equation}
p(\mathbf{r}_n | s_{\le n}, \mathbf{r}_{< n}) = 
\begin{cases}
     \frac{N_{s_n}^2}{2\pi(1+N_{s_n})} \frac{\xi_{s_n}^{-1}} { [1 +\Delta(\mathbf{r}_n)/\xi_{s_n}]^{1+N_{s_n}} } & \!\!\text{if} \; N_{s_n}\! \geq \!1\\
    1              & \!\!\text{if} \; N_{s_n}\! = \!0
    \nonumber
\end{cases}
\end{equation}
where $N_{s_n} =  \sum_{i=1}^{n-1} \mathbb{I}[s_i=s_n]$ is the number of posts assigned to spatiotemporal pattern $s_n$, 
$\xi_{s_n}$ is given by\footnote{Up to a factor of $\tfrac{1}{2}N_{s_n}$ this is the spatial variance of pattern $s_n$ when $\beta_{\rm space} \to 0$.}
\begin{equation}
\nonumber
\xi_{s_n} = \beta_{\rm space} +  \tfrac{1}{2}\sum_{i=1}^{n-1}\mathbf{r}_i^2 \mathbb{I}[s_i=s_n] -\tfrac{1}{2N_{s_n}}\left( \sum_{i=1}^{n-1}\mathbf{r}_i \mathbb{I}[s_i=s_n] \right)^2 
\end{equation}
\vspace{-3mm}
\!and
\begin{equation}
\Delta(\mathbf{r}_n) = \frac{N_{s_n}}{2(N_{s_n}+1)}\left(\mathbf{r}_n - \frac{1}{N_{s_n}}\sum_{i=1}^{n-1}\mathbf{r}_i\mathbb{I}[s_i=s_n]\right)^2 
\nonumber
\end{equation}
This choice of $q_n(\cdot)$ results in the incremental importance weight
\begin{equation}
\label{eqn:incrW}
\alpha_n(s_{< n}) = p(t_n | s_{<n}, t_{<n}) Q_n(s_{<n},  t_{\le n}, \mathbf{d}_{\le n}, \mathbf{r}_{\le n})
\end{equation}
where $Q_n(s_{<n},  t_{\le n}, \mathbf{d}_{\le n}, \mathbf{r}_{\le n})$ is given by
\begin{equation}
\sum_{s_n} p(s_n | s_{<n}, t_{\le n}) p(\mathbf{d}_n | s_{\le n}, \mathbf{d}_{<n}) p(\mathbf{r}_n | s_{\le n}, \mathbf{r}_{<n})
\end{equation}
This update is optimal in the sense that it leads to minimum variance among the particle weights. 
Finally note that in order to mitigate against particle degeneracy systematic resampling is used whenever the particle system
satisfies $\|\textrm{\bf w}_n\|_2^{-2}< \kappa_{\rm thresh} |\Pcal| $ (throughout we use $\kappa_{\rm thresh}=0.9$). 
For more details on Sequential Importance Resampling see e.g.~ref.~\cite{doucet2009tutorial}.
%

\begin{algorithm}[t]
\DontPrintSemicolon 
\small
  Initialize $w^{(p)}_1\to 1/ |\Pcal |$ and  $S^{(p)} \to 0$ for all $p \in \Pcal$. \\
  \For{$n=1,\ldots,N$}{
  \For{$p \in \Pcal$}{
    Draw $s^{(p)}_n$ from~Eqn.~\ref{eqn:proposal_distribution}.

    \If {$s^{(p)}_n =S^{(p)}+1$}{
     Draw the time kernel parameters $\{ \alpha_s, \tau_s \}$ for $s={s^{(p)}_n}$ from the prior\; \\
     Increase the number of patterns $S^{(p)} \to S^{(p)}+1$\;
     }
     Update the particle weight $\textrm{w}^{(p)}_n$ using Eqn.~\ref{eqn:incrW}\;
     Update $\{ \alpha_s, \tau_s \}$ for all patterns via Eqn.~\ref{eqn:alpha_mle}\;
   }
    Normalize particle weights. \\
     \If {$\|\textrm{\bf w}_n\|_2^{-2}< \kappa_{\rm thresh} |\Pcal| $}{
     Resample particles.
     }
 }
 Finally return the particle $p \in \Pcal$ with the largest weight as an approximate MAP estimate to Eqn.~\ref{eqn:map}.
  \caption{Inference algorithm for the SDHP\label{alg:inference}}
\end{algorithm}

\subsubsection{Time kernel inference}
If the time kernels parameters, $\{ \alpha_{s}, \tau_{s} \}$, are fixed the inference procedure described above yields an unbiased estimate of the 
posterior $p(s_{1:N} | t_{1:N}, \mathbf{d}_{1:N}, \mathbf{r}_{1:N})$. 
In general, however, these parameters are unknown and need to be estimated.
Methods for calculating the full posterior $p(s_{1:N}, \{ \alpha_{s}, \tau_{s} \} | t_{1:N}, \mathbf{d}_{1:N}, \mathbf{r}_{1:N})$ can be derived; however, they are computationally expensive and do not scale to large datasets (since they rely on e.g.~expensive MCMC updates). Since our primary interest is not in the posterior itself but rather the MAP estimate, i.e.
\begin{equation}
\label{eqn:map}
s_{1:N}^{\rm MAP} = \underset{s_{1:N}}{\arg\max} \; p(s_{1:N} | t_{1:N}, \mathbf{d}_{1:N}, \mathbf{r}_{1:N}),
\end{equation}
we do not necessarily require SMC to produce unbiased samples from the posterior. Rather, we just need SMC to explore the posterior space efficiently and 
return an (approximate) MAP estimate.
Consequently, we use the following computationally efficient procedure: 
after each time step, the parameters $\{ \alpha_{s}, \tau_{s} \}$ are set equal to a (restricted) MLE estimate. More specifically, as part of the model specification
we choose a fixed, finite set of allowed time constants, $ \Psi_\tau = \{ \tau_i \} $. 
Then at each time step $n$ and for each spatiotemporal pattern $s$ and $\tau_i \in \Psi_\tau$ we compute 
\vspace{.2cm}
\begin{equation}
\label{eqn:alpha_mle}
\alpha_s^{\rm MLE}(\tau_i) = \underset{\alpha_s}{\arg\max} \; p(\alpha_s | \alpha_{\rm time}, \beta_{\rm time})
 p(\mathcal{T}_{s; n} | \alpha_s, \tau_i)
\end{equation}
where $\mathcal{T}_{s; n}$ is the sequence of times for the posts assigned to spatiotemporal pattern $s$ through time step $n$. For each 
$\tau_i \in \Psi_\tau$ Eqn.~\ref{eqn:alpha_mle} can be computed in closed form. 
Finally, we choose the pair $(\alpha_s^{\rm MLE}(\tau_i), \tau_i)$ that maximizes the likelihood in Eqn.~\ref{eqn:alpha_mle}.\footnote{Note that this 
is not equivalent to simultaneously maximizing over $(\alpha_s, \tau_s)$, which cannot be done in closed form.} In this way the parameters $\{ \alpha_{s}, \tau_{s} \}$ are updated at each time step for all patterns that contain at least two posts.

\subsection{Setup for synthetic experiments}
Unless stated otherwise, the following experimental parameters are common to all four experiments: 
the vocabulary has length $| \Vcal | =15$;
the hyperparameters for the prior on the self-excitation parameter $\alpha_s$ are given by $\alpha_{\rm time} = 0.1$ and $\beta_{\rm time} = 0.2$;
the base intensity $\lambda_0=10$;
the time constants are given by $\Psi_\tau = \{ 1 \}$;
the Dirichlet hyperparameter is given by $\theta_0 = 1$;
the number of words per tweet is given by $N_{\rm words}=7$;
and the number of particles used during inference is $| \Pcal | = 4$. 
The number of tweets in each sample will be denoted as $N$ and 
the number of trials per value of $x$ will be denoted as $N_{\rm trials}$.

In order for the spatial part of the generative process to be well-defined, we use a uniform prior on the 
mean location $\mathbf{R}_s$ of each pattern $s$, with the prior defined on the unit square.\footnote{If a given tweet falls
outside the unit square during sampling from the generative process, sampling of the location is repeated until the location falls
within the unit square.} Unless stated otherwise
the spatial hyperparameter $\beta_{\rm space} = 0.01$ and the generative process assigns each spatiotemporal pattern
a spatial extent $\sigma_0 = 0.1$.

For the experiment corresponding to Fig.~2a we set $N_{\rm trials}=60$
and $\sigma_0 = 0.03$, while for the experiment corresponding to Fig.~2b we set $N=5500$ and $N_{\rm trials}=500$ as well as 
$N_{\rm words}=15$ and $\sigma_0 = 0.02$ (so that even smaller patterns should be readily identifiable).
For the experiment corresponding to Fig.~3a we set $N=500$, $N_{\rm trials}=50$, 
$|\Pcal|=8$, and $\beta_{\rm space} = \sigma_0^2$, 
while for the experiment corresponding to Fig.~3b we set $N=2000$, $N_{\rm trials}=200$, $\sigma_0 = 0.03$, and $|\Pcal|=1$.

\subsection{Details on Location Prediction Experiment}
The two selection criteria used in the paper are defined as follows:

\emph{--- Loose selection:} we sort all tweets in ascending order according to the $\sigma_s$ of the associated pattern,
discard any tweet in a pattern with less than $7$ tweets, and compute the average root mean square error (RMSE) of the top 4\%.

\emph{--- Tight selection:} we sort all tweets in ascending order according to the $\sigma_s$ of the associated pattern,
discard any tweet in a pattern with less than $11$ tweets, and compute the average root mean square error (RMSE) of the top 4\%.

In the above measures, ties are adjudicated by preferring tweets which belong to patterns with more tweets. Any remaining ties 
are decided randomly. 

\subsection{Goodness of Fit Measures}
\subsubsection{Spatial measure}
We use the following spatial goodness of fit measure. At each iteration $n$ of the corresponding inference algorithm (after a burnin period of 500 tweets), 
we evaluate the marginal likelihood of the next sample $n+1$ given the parameters and latent variables inferred from the first $n$ samples; e.g.~for the SDHP we have:
\begin{equation}
\label{eqn:spatialgof}
\!{\rm spatial \; g.o.f.} = \tfrac{1}{2000}\!\!\sum_{n=501}^{2500}\!\! \log p(\rb_n |  t_{\le n},  s_{<n},  \db_{\le n},  \rb_{<n})  
\end{equation}
An analogous expression (i.e.~without conditioning on $\db_{\le n}$ and $t_{\le n}$) holds for the GMM. 
In order to make a more direct comparison between the two models we setup the GMM as follows:
(i) at each iteration $n$ we set the number of gaussian components equal to the number of spatiotemporal patterns inferred by the SDHP at time step $n-1$; and
(ii) we consider isotropic gaussians with the minimum covariance set equal to ${\sigma^2}_{\rm{min}} = 2 \beta_{\rm{space}}$.
\subsubsection{Content measure}
With reference to the expression in Eqn.~\ref{eqn:spatialgof}, we use a related goodness of fit measure, namely the perplexity $\Pcal$~\cite{LDA}; 
e.g.~for the DHP we have the following:
\begin{equation}
\Pcal \! = \exp \!\left(\!\! \tfrac{-1}{N_{\rm{words}}} \!\!\sum_{n=501}^{2500} \!\!\log p(\db_n |  t_{\le n},  s_{<n},  \db_{<n}) \!\right) 
\nonumber
\end{equation} 
where $N_{\rm{words}}$ is the total number of words in $\{\db_{501}, ..., \db_{2500} \}$.
An analogous expression (i.e.~with additional conditioning on $\rb_{\le n}$) holds for the SDHP.
In order to make a more direct comparison between the two models we set $\lambda_0$ for the DHP such that the number of inferred
patterns matches that of the SDHP.

\end{document}